\documentclass [12pt]{article}
\usepackage{amsmath,amssymb,cite}
\usepackage{appendix}
\usepackage{feynmp}
\usepackage[vcentermath,enableskew]{youngtab}
\usepackage{tabularx}
\usepackage[english]{babel}
\usepackage[dvips]{graphicx}
\usepackage{indentfirst}
\usepackage{epsfig}
\usepackage{physics}
\usepackage{pdfpages}
\usepackage{cases}
\usepackage{appendix}
\usepackage{tabularx}
\usepackage[english]{babel}
\usepackage[dvips]{graphicx}
\usepackage{indentfirst}
\usepackage{epsfig}
\usepackage{slashed}
\usepackage{fancyhdr}
\usepackage{cases}
\usepackage{tcolorbox}
\usepackage{amsmath,amssymb,cite}
\usepackage{amsfonts, geometry, hyperref}
\usepackage{booktabs,longtable}
\usepackage{appendix}
\usepackage{feynmp}
\usepackage{float}
\usepackage{subfigure}
\usepackage[vcentermath,enableskew]{youngtab}
\usepackage{tabularx}
\usepackage[english]{babel}
\usepackage{graphicx}
\usepackage{indentfirst}
\usepackage{epsfig}
\usepackage{color}
\usepackage{epstopdf}
\usepackage[]{hyperref}
\usepackage[section]{placeins}
\usepackage[stable]{footmisc}
\usepackage{appendix}
\usepackage{adjustbox}
\usepackage{tabularx}
\usepackage[english]{babel}
\usepackage{graphicx}
\usepackage{indentfirst}
\usepackage{epsfig}
\usepackage{slashed}
\usepackage{fancyhdr}
\usepackage{url}
\usepackage{pgfplots}
\usepackage{amsmath}

\pgfplotsset{compat=1.18}
\numberwithin{equation}{section}

\makeatletter
\patchcmd{\beamer@sectionintoc}
  {\vfill}
  {\vskip\itemsep}
  {}
  {}

\long\def\beamer@section[#1]#2{%
  \beamer@savemode%
  \mode<all>%
  \ifbeamer@inlecture
    \refstepcounter{section}%
    \beamer@ifempty{#2}%
    {\long\def\secname{#1}\long\def\lastsection{#1}}%
    {\global\advance\beamer@tocsectionnumber by 1\relax%
      \long\def\secname{#2}%
      \long\def\lastsection{#1}%
      \addtocontents{toc}{\protect\beamer@sectionintoc{\the\c@section}{#2\hfill\the\c@page}{\the\c@page}{\the\c@part}%
        {\the\beamer@tocsectionnumber}}}%
    {\let\\=\relax\xdef\sectionlink{{Navigation\the\c@page}{\noexpand\secname}}}%
    \beamer@tempcount=\c@page\advance\beamer@tempcount by -1%
    \beamer@ifempty{#1}{}{%
      \addtocontents{nav}{\protect\headcommand{\protect\sectionentry{\the\c@section}{#1}{\the\c@page}{\secname}{\the\c@part}}}%
      \addtocontents{nav}{\protect\headcommand{\protect\beamer@sectionpages{\the\beamer@sectionstartpage}{\the\beamer@tempcount}}}%
      \addtocontents{nav}{\protect\headcommand{\protect\beamer@subsectionpages{\the\beamer@subsectionstartpage}{\the\beamer@tempcount}}}%
    }%
    \beamer@sectionstartpage=\c@page%
    \beamer@subsectionstartpage=\c@page%
    \def\insertsection{\expandafter\hyperlink\sectionlink}%
    \def\insertsubsection{}%
    \def\insertsubsubsection{}%
    \def\insertsectionhead{\hyperlink{Navigation\the\c@page}{#1}}%
    \def\insertsubsectionhead{}%
    \def\insertsubsubsectionhead{}%
    \def\lastsubsection{}%
    \Hy@writebookmark{\the\c@section}{\secname}{Outline\the\c@part.\the\c@section}{2}{toc}%
    \hyper@anchorstart{Outline\the\c@part.\the\c@section}\hyper@anchorend%
    \beamer@ifempty{#2}{\beamer@atbeginsections}{\beamer@atbeginsection}%
  \fi%
  \beamer@resumemode}%

\def\beamer@subsection[#1]#2{%
  \beamer@savemode%
  \mode<all>%
  \ifbeamer@inlecture%
    \refstepcounter{subsection}%
    \beamer@ifempty{#2}{\long\def\subsecname{#1}\long\def\lastsubsection{#1}}
    {%
      \long\def\subsecname{#2}%
      \long\def\lastsubsection{#1}%
      \addtocontents{toc}{\protect\beamer@subsectionintoc{\the\c@section}{\the\c@subsection}{#2\hfill\the\c@page}{\the\c@page}{\the\c@part}{\the\beamer@tocsectionnumber}}%
    }%
    \beamer@tempcount=\c@page\advance\beamer@tempcount by -1%
    \addtocontents{nav}{%
      \protect\headcommand{\protect\beamer@subsectionentry{\the\c@part}{\the\c@section}{\the\c@subsection}{\the\c@page}{\lastsubsection}}%
      \protect\headcommand{\protect\beamer@subsectionpages{\the\beamer@subsectionstartpage}{\the\beamer@tempcount}}%
    }%
    \beamer@subsectionstartpage=\c@page%
    \edef\subsectionlink{{Navigation\the\c@page}{\noexpand\subsecname}}%
    \def\insertsubsection{\expandafter\hyperlink\subsectionlink}%
    \def\insertsubsubsection{}%
    \def\insertsubsectionhead{\hyperlink{Navigation\the\c@page}{#1}}%
    \def\insertsubsubsectionhead{}%
    \Hy@writebookmark{\the\c@subsection}{#2}{Outline\the\c@part.\the\c@section.\the\c@subsection.\the\c@page}{3}{toc}%
    \hyper@anchorstart{Outline\the\c@part.\the\c@section.\the\c@subsection.\the\c@page}\hyper@anchorend%
    \beamer@ifempty{#2}{\beamer@atbeginsubsections}{\beamer@atbeginsubsection}%
  \fi%
  \beamer@resumemode}

\makeatother

\begin{document}

\title{Molien--Weyl Singlet Counting and BFSS$_2$--Factorization in Gaussian Matrix Quantum Mechanics\\
}

\author{
Badis Ydri\\[2mm]
Department of Physics, Badji Mokhtar Annaba University, Algeria\\
}

\maketitle

\begin{abstract}

We study the singlet-sector structure of mass-deformed BFSS$_{d+1}$ matrix quantum mechanics by combining the large--\(d\) Gaussian reduction with the Molien--Weyl projection. The Gaussian reduction captures the bulk matrix dynamics through a gauged harmonic oscillator, while the Molien--Weyl integral imposes the Gauss law and reorganizes the physical Hilbert space into holonomy-projected singlet excitations.

We show that the very-low-temperature bosonic singlet spectrum is universally controlled by the quadratic Gram operators \(\Tr(X_aX_b)\), whose number is \(d(d+1)/2\). For \(N=2\), this result is established by explicit residue computations and character methods; for \(N>2\), it is supported by the character analysis. Thus the infrared spectrum begins as a collection of BFSS$_2$--like Gram towers, although higher invariant structures generally modify the full partition function.

We also give a Hamiltonian derivation of the exceptional exact factorization at \((d,N)=(2,2)\), where the BFSS$_3$ singlet partition function equals the cube of the BFSS$_2$ one for all temperatures. This rigidity is special to the \(SU(2)\) invariant tensor structure and distinguishes exact singlet factorization from the possibility of smooth deconfinement--like holonomy crossovers at finite $N$. Finally, we extend the Gram-counting picture to supersymmetric BFSS/BMN models and indicate how the Molien--Weyl formulation can benchmark Monte Carlo simulations in both \(X_a\)-space and holonomy space.
\end{abstract}


\newpage
\tableofcontents

\newpage
\section{Introduction, goal, and summary}

\medskip
\noindent
\subsection{Generalities: BFSS/BMN systems}

\medskip
\noindent
A basic class of matrix quantum mechanical systems is obtained by dimensionally reducing
supersymmetric Yang--Mills theory to one time dimension~\cite{Brink:1976bc}. Starting from ten-dimensional
\(\mathcal N=1\) super Yang--Mills theory, one obtains gauged quantum
mechanics of adjoint matrix coordinates \(X_a\), \(a=1,\ldots,d\), with Euclidean bosonic action
\begin{eqnarray}
S_{\rm BFSS,B}^{\rm E}
=
\frac{1}{g^2}
\int_{0}^{\beta}\! dt~ 
\Tr\!\left[
\frac{1}{2}(D_t X_a)^2
-
\frac{1}{4}[X_a,X_b]^2
\right]
+
\text{fermionic terms},
\qquad
D_t=\partial_t-i[A_t,\cdot].
\label{BFSSIntro}
\end{eqnarray}
The allowed supersymmetric dimensions are constrained by the Fierz identities of
Baake, Reinicke and Rittenberg~\cite{Baake:1984ie}, giving
\[
D_{\rm YM}=d+1=10,6,4,3,2,
\qquad
D_{\rm M}=d+2=11,7,5,4,3.
\]
The planar or holographic regime is the usual 't~Hooft limit~\cite{tHooft1974},
with \(N\to\infty\), \(g^2\to0\), and \(\lambda=g^2N\) fixed. This is the natural
large-\(N\) setting underlying holography~\cite{tHooft1993,Susskind1995}.

\medskip
\noindent
The most important member of this family is the \(d=9\) BFSS$_{10}\) model, or
M-(atrix) theory~\cite{BanksFischlerShenkerSusskind1997}. It describes the
low-energy worldvolume dynamics of \(N\) coincident D\(0\)-branes~\cite{Witten1996},
whose type IIA supergravity dual is the black 0-brane background~\cite{Itzhaki1998}.
D\(0\)-branes arise naturally in type IIA string theory~\cite{Polchinski1995}, while
type IIA supergravity itself follows from compactifying eleven-dimensional
supergravity~\cite{Cremmer1978,Witten1995}. Correspondingly, the M-wave solution
reduced along the compact eleventh direction gives the black 0-brane geometry
\cite{Hyakutake:2014maa,Hyakutake:2006aq}. This connects the BFSS model to the
matrix regularization of the light-cone supermembrane
\cite{Hoppe1982,Hoppe1988,deWitHoppeNicolai1988} and to light-cone superparticles
in maximally supersymmetric pp-wave backgrounds
\cite{Kowalski-Glikman:1984qtj,Blau:2001ne}.

\medskip
\noindent
In the D\(0\)-brane interpretation, the diagonal entries of the matrices \(X_a\)
represent brane positions, whereas the off-diagonal entries describe open strings
stretched between distinct branes. When the branes coincide, these off-diagonal
modes become light and produce the non-Abelian gauge dynamics \cite{Azeyanagi2009}. For a pedagogical 
presentation of this picture see~\cite{Zwiebach2009,Becker2006}.
Maldacena's gauge/gravity correspondence~\cite{Maldacena1999,Gubser1998,Witten1998}
then relates the strongly coupled large-\(N\) gauge theory to weakly curved type II
string theory in the appropriate black 0-brane background. Since the gauge theory
admits a nonperturbative lattice definition~\cite{Wilson:1974sk}, this provides a
concrete framework for studying quantum gravity and black-hole thermodynamics.
This duality has been tested extensively by Monte Carlo simulations
\cite{Catterall2008,Anagnostopoulos2008,Hanada2014,Hanada2016b,Filev:2015hiaF}
and by analytic methods~\cite{Kabat2001,Hanada2009,Hyakutake2014}; see also the
review~\cite{Hanada2016}.

\medskip
\noindent
The lower-dimensional BFSS$_{d+1}$ models, with \(D_{\rm YM}=d+1=6,4,3,2\), may be
viewed as reduced analogues of BFSS$_{10}$. They retain many of the same structural
features---gauge dynamics, singlet constraints, confinement/deconfinement behavior,
eigenvalue condensation, and emergent geometry---while being more accessible both
analytically and numerically. They can also be interpreted as lower-dimensional
light-cone supermembrane or superparticle matrix systems in the corresponding
dimensions \(D_{\rm M}=d+2\).

\medskip
\noindent
A second important class is obtained by adding supersymmetric mass deformations.
The prototype is the BMN plane-wave matrix model~\cite{BerensteinMaldacenaNastase2002},
and the general massive supersymmetric Yang--Mills quantum mechanical deformations
were classified in~\cite{Kim:2006,Park:2005}. These deformations add quadratic mass
terms, Myers-type cubic couplings~\cite{Myers}, and fermionic mass terms in such a
way that the required supersymmetries are preserved. Schematically,
\begin{eqnarray}
S_{\rm BMN,B}^{\rm E}
=
S_{\rm BFSS,B}^{\rm E}
+
\frac{1}{g^2}
\int_{0}^{\beta}\! dt~
\Tr\!\left[
\mu_1 X_a^2+\mu_2\epsilon_{ijk}X_iX_jX_k
\right]
+
\text{fermionic terms}.
\label{BMNIntro}
\end{eqnarray}
The allowed BMN models include BMN$_{3,4,6,10}\), corresponding to \(d=2,3,5,9\),
while the special \(d=1\) model, BMN$_2\), is discussed in~\cite{Park:2005,Ydri2025}.
The mass terms lift the flat directions of BFSS and introduce the curvature scale
of the pp-wave background; in the maximally supersymmetric case, the corresponding
half-BPS sectors are related to LLM bubbling geometries~\cite{Lin:2004nb}. Monte
Carlo studies of BMN-type models include~\cite{Asano:2018nol,Asano:2020yry}.

\medskip
\noindent  In summary, each BFSS\(_{d+1}\) model admits a corresponding BMN deformation preserving maximal supersymmetry. These deformations describe supermembranes and superparticles in maximally supersymmetric pp-wave backgrounds. The corresponding classification is summarized in Table~\ref{so3}.

\begin{table}[h]
\centering
\begin{tabular}{@{}lllll@{}}
\toprule
\textbf{Model} & \(D_{\rm YM}\) & Splitting of \(SO(D_{\rm YM}-1)\) & Superalgebra & Deformation parameter \\
\midrule
\(\mathcal N=16\) & 10 & \(SO(6)\times SO(3)\) & \(\mathfrak{su}(2|4)\) & \(\mu\) \\
\(\mathcal N=8\) type I & 6 & \(SO(3)\times SO(2)\) & \(\mathfrak{su}(2|2)\) & \(\mu\) \\
\(\mathcal N=8\) type II & 6 & \(SO(4)\) & \(\mathfrak{su}(2|1)\oplus\mathfrak{su}(2|1)\) & \(\mu\) \\
\(\mathcal N=4\) type I & 4 & \(SO(3)\) & \(\mathfrak{su}(2|1)\) & \(\mu_1,\mu_2\) \\
\(\mathcal N=4\) type II & 4 & \(SO(2)\) & \(\mathrm{Clifford}_{4}(\mathbb R)\) & \(\mu\) \\
\(\mathcal N=2\) & 3 & \(SO(2)\) & \(\mathrm{Clifford}_{2}(\mathbb R)\) & \(\mu\) \\
\(\mathcal N=1+1\) & 2 & \(SO(1,2)\) & \(\mathfrak{osp}(1|2,\mathbb R)\) & \(\Lambda(t),\rho(t)\) \\
\bottomrule
\end{tabular}
\caption{Classification of massive supersymmetric Yang--Mills quantum mechanics models and their deformation parameters.}
\label{so3}
\end{table}

\medskip
\noindent
A third regime, central to the present work, is the Gaussian or large-mass
approximation to BFSS/BMN matrix quantum mechanics. In this limit the theory
reduces to a supersymmetric gauged matrix harmonic oscillator, whose singlet
partition function can be written as a Molien--Weyl integral
\cite{OConnor:2023mss,OConnor:2024udv}. Mathematically, these integrals compute
Hilbert series of invariant operators~\cite{CoxLittleOShea2005}. The same Gaussian
structure also emerges dynamically in the large-\(d\) expansion of BFSS models, where
the Yang--Mills interaction is replaced by an effective mass scaling as
\(s\sim d^{1/3}\)~\cite{Mandal:2009vzN,Mandal:2011hbN,Kabat:2000zv,Kabat:2001ve}.
For BMN systems, a correlated large-mass/large-\(d\) double-scaling limit was
identified in~\cite{Ydri2025}. This Gaussian/Molien--Weyl regime is the starting
point for the singlet-sector analysis developed below.

\subsection{Goal: towards matrix quantum mechanics and matrix quantum gravity}

\medskip
\noindent
This work, which consists of several parts, concerns gauge theory in one dimension with an arbitrary number \(d\) of noncommuting coordinate matrices. More precisely, we study the celebrated BFSS$_{d+1}$ matrix quantum mechanics and its mass--deformed BMN$_{d+1}$ extensions, employing a combination of large--\(d\) Gaussian reduction, Molien--Weyl singlet projection, and Monte Carlo methods in order to probe the structure of matrix quantum mechanics and its possible matrix quantum gravity interpretations.

\medskip
\noindent
Here, quantum gravity may refer either to the gauge/gravity duality approach and its quantum black-hole dynamics, or to the noncommutative-geometry/matrix-model approach to emergent geometry and gravity \cite{Ydri:2022ueu,Ydri:2021cam,Ydri:2020fry}.

\medskip
\noindent
The present work rests on two working assumptions. The first is that matrix quantum mechanics provides the more fundamental dynamical framework, while zero-dimensional matrix models, including the seminal IKKT model~\cite{Ishibashi:1996xs}, may be understood, from this perspective, as reductions, limits, or approximations of a genuinely one-dimensional matrix quantum theory. The second assumption is what one may call the ``unreasonable effectiveness'' of the Gaussian structure of matrix quantum mechanics, namely the assumption that this structure is already sufficiently rich to capture a significant part of the nontrivial quantum physics of the full theory.

\medskip
\noindent 
The broader project is organized around the following themes \cite{Ydri2025}:
\begin{enumerate}
\item Revisiting the large--\(d\) limit of the BFSS/BMN system.

\item The Molien--Weyl integral, singlet counting, and BFSS$_2$ factorization.

\item The endpoint formulation of \(N=2\) Gaussian BFSS/BMN matrix quantum mechanics, Wishart/Stiefel entropy, and emergent Grassmannian geometry.

\item Matrix quantum gravity and Monte Carlo simulations of the supersymmetric BFSS$_3$/BMN$_3$ system.

\item BFSS$_2$/BMN$_2$ quantum gravity, AdS$_2$/dS$_2$ noncommutative geometry, and emergent/latent geometry.
\end{enumerate}

\medskip
\noindent
The main emphasis of the present paper, which is the second in the series, is the Molien--Weyl integral, primarily at \(N=2\), its use in counting singlet states, and our observation of BFSS$_2$ factorization within higher BFSS$_{d+1}$ models.

\subsection{Organization of the paper}

\medskip
\noindent
The paper is organized as follows. In Section~\ref{section2} we review the large--\(d\) Gaussian
reduction of the mass--deformed BFSS$_{d+1}$ matrix model and derive the
Molien--Weyl holonomy formulation from the gauged matrix harmonic oscillator.
Section~\ref{section3} develops the bosonic Molien--Weyl counting of singlet states, with
emphasis on the universal Gram--operator contribution and the emergence of
BFSS$_2$--like singlet towers. Section~\ref{section4} extends the Molien--Weyl construction to
supersymmetric BFSS/BMN models and derives the corresponding universal
quadratic singlet counting law.

\medskip
\noindent
In Section~\ref{section5} we give a Hamiltonian derivation of the exact BFSS$_2$--factorization
of the \(N=2\) BFSS$_3$ model, and clarify its relation to finite--\(N\)
deconfinement and the exceptional role of \(N=2\). Section~\ref{section6} discusses our Monte
Carlo simulation strategy for BFSS/BMN systems, both in the Gaussian
approximation and beyond, including full, bosonic, and Molien--Weyl--based
effective descriptions.

\medskip
\noindent
Further technical details, including explicit residue
calculations and illustrative Monte Carlo results,
are available in the original arXiv version of
this work~\cite{Ydri:2026MolienV1}.

\section{Large--$d$ and Molien--Weyl integrals for BFSS/BMN systems}\label{section2}
\subsection{BFSS$_{d+1}$ model at large $d$}

One of our basic theories here is the Euclidean BFSS$_{d+1}$ Yang-Mills matrix quantum mechanics defined by the action

\begin{eqnarray}
S
&=&
N\int_{0}^{\beta}\! dt\ \mathrm{Tr}\bigg[\frac{1}{2}(D_tX_a)^2
+\frac{m}{2}X_a^2
-\frac{1}{4}[X_a,X_b]^2\bigg].
\end{eqnarray}
We have analyzed in \cite{Ydri2025} the large--$d$ dynamics of this model in a correlated double--scaling limit in
which both the mass parameter $m$ and the number of matrices $d$ are taken large while the
combination
\begin{eqnarray}
\kappa^{2/3}
\;\equiv\;
\frac{m}{d^{2/3}}
\end{eqnarray}
is held fixed.

In this limit, the large--$d$ expansion renders the original commutator--squared interaction
self--consistently Gaussian, effectively replacing it by a dynamical mass term.
The resulting effective action is that of a gauged matrix harmonic oscillator,
\begin{eqnarray}
S_{\rm MHO}[X;\theta]
&=&
N\int_{0}^{\beta}\! dt\ \mathrm{Tr}\bigg[
\frac{1}{2}(D_tX_a)^2
+\frac{s^2}{2}X_a^2
\bigg],\label{MHO_action_for_Ward}
\end{eqnarray}
where the effective mass is
\begin{eqnarray}
s^2 = m + k_0,
\end{eqnarray}
with $k_0$ determined self--consistently by the gap equation:
\begin{eqnarray}
s^3-m\,s=d,
\qquad d>0.
\end{eqnarray}
The corresponding confinement/deconfinement transition is governed by the holonomy effective
action \cite{Kawahara:2007fnF,Aharony:2003sxF,Aharony:2004igF,AlvarezGaume:2005fvF, Gross:1980heF,Wadia:1980cpF}, yielding a critical temperature
\begin{eqnarray}
T_c(\kappa)
=
\frac{s(\kappa)}{\log d}
=
\frac{d^{1/3}}{\log d}
\Big(\kappa^{1/3}+\frac{1}{2\kappa^{2/3}}+\cdots\Big),
\end{eqnarray}
which is parametrically pushed to higher values as $d\to\infty$.

\medskip
\noindent
For \(\mathrm{BFSS}_2\), equation \eqref{MHO_action_for_Ward} is the exact gauged matrix quantum mechanics action, not a Gaussian approximation. In this case there is a single adjoint matrix, so the label \(a\) should be understood as taking only one value.

Our main conclusions in \cite{Ydri2025} are as follows:
\begin{itemize}
\item
In the correlated double--scaling limit, the critical temperature grows parametrically with $d$,
so that for any fixed physical temperature the theory is generically dominated by the
\emph{uniform} (confining) holonomy phase.

\item
There exists an overlap window in which both the low–temperature and high–temperature saddle–point analyses are simultaneously reliable. In this window, explicit
evaluation of observables shows that noncommutativity is parametrically suppressed, and the
matrices are driven to approximately commuting configurations. The resulting dynamics is therefore
IKKT--like in the Yang-Mills phase: spacetime emerges from an almost--commutative matrix geometry, with the remaining
degrees of freedom encoded in a holonomy--dominated, effectively zero--dimensional theory.
\end{itemize}

\medskip
\noindent
The large--$d$ Gaussian reduction describes $d$ adjoint
matrix oscillators with a common self--consistent
frequency $s$. At fixed $s$, their leading Gaussian
contributions to the extent of space scale linearly
with $d$, as discussed in \cite{Ydri2025}.
This bulk decomposition, however, should not be
confused with a factorization of the physical
singlet Hilbert space. Imposing the Gauss law
through the Molien--Weyl projection reorganizes
the oscillator excitations into gauge--invariant
states. As we show in Section~\ref{section3},
the leading low--temperature singlet contribution
is controlled by the $d(d+1)/2$ quadratic Gram
operators $\Tr(X_aX_b)$, rather than by $d$
independent singlet towers.

\subsection{Molien--Weyl reduction for $\mathrm{BFSS}_2$ model}
\subsubsection{Lattice gauged Laplacian}

\medskip
\noindent
In this subsection we derive the Molien--Weyl integral for the bosonic $\mathrm{BFSS}_2$ model,
i.e. a single adjoint matrix harmonic oscillator coupled to a $U(N)$ gauge field on the thermal
circle \cite{OConnor:2023mss,OConnor:2024udv}. The Euclidean action is \cite{Park:2005}
\begin{eqnarray}
S_{\rm MHO}[X,A_0]
&=&
N\int_0^\beta dt\ \Tr\bigg[
\frac{1}{2}(D_t X)^2-\frac{\Lambda(t)}{2}X^2-\rho(t)X
\bigg],
\qquad
D_t X=\partial_t X-i[A_0,X],\nonumber\\
&&-\Lambda(t)\equiv s^2,\qquad -\rho(t)\equiv 0.\label{eq:MHO_action}
\end{eqnarray}
Here, $X(t)$ is Hermitian and traceless (adjoint of $SU(N)$), and $s>0$ is the oscillator frequency.
The partition function is
\begin{eqnarray}
Z_{\rm MHO}(\beta,s)
=
\int {\cal D}A_0\,{\cal D}X\ e^{-S_{\rm MHO}[X,A_0]}.
\label{eq:Z_def}
\end{eqnarray}

\medskip
\noindent We discretize the thermal circle into $\Lambda$ sites with spacing $a=\beta/\Lambda$ and denote
$X_\ell\equiv X(t_\ell)$ with $t_\ell=\ell a$. Introduce link variables $U_{\ell,\ell+1}\in U(N)$ implementing parallel transport between adjacent
time-slices, so that the covariant difference is
\begin{eqnarray}
D_t X(t)\ \longrightarrow\ \frac{1}{a}\Big(U_{\ell,\ell+1}\,X_{\ell+1}\,U_{\ell,\ell+1}^{-1}-X_\ell\Big).
\label{eq:lattice_covdiff}
\end{eqnarray}
The gauged lattice action then reads
\begin{eqnarray}
S_{\Lambda}[X,U]
=
\frac{N}{2a}\sum_{\ell=0}^{\Lambda-1}\Tr\Big(U_{\ell,\ell+1}\,X_{\ell+1}\,U_{\ell,\ell+1}^{-1}-X_\ell\Big)^2
+\frac{Na s^2}{2}\sum_{\ell=0}^{\Lambda-1}\Tr X_\ell^2.
\label{eq:lattice_action}
\end{eqnarray}
The theory is invariant under local gauge transformations
$X_\ell\to \Omega_\ell X_\ell\Omega_\ell^{-1}$,
$U_{\ell,\ell+1}\to \Omega_\ell\,U_{\ell,\ell+1}\,\Omega_{\ell+1}^{-1}$ with $\Omega_\ell\in U(N)$.
Using this invariance (together with Haar invariance of the $U_{\ell,\ell+1}$ measures), one can
gauge--transform all links to unity except the closing link, so that the entire gauge dependence is
captured by a single holonomy \cite{Filev:2015hiaF}
\begin{eqnarray}
g\ \equiv\ \prod_{\ell=0}^{\Lambda-1}U_{\ell,\ell+1}\ \in\ U(N).
\label{eq:holonomy_def}
\end{eqnarray}
Equivalently, one may fix $U_{\ell,\ell+1}=\mathbf{1}$ for $\ell=0,\dots,\Lambda-2$ and
$U_{\Lambda-1,0}=g$.

\subsubsection{Gauged quadratic kernel and holonomy determinant}

\medskip
\noindent
We start from the compact operator form of the gauged lattice Laplacian
\begin{eqnarray}
a^2\Delta_{\Lambda,g}
\;\equiv\;
(1-e^{aD_\tau})(1-e^{-aD_\tau})
\;=\;
2-e^{aD_\tau}-e^{-aD_\tau},
\label{eq:Delta_op}
\end{eqnarray}
where the covariant shift operators are defined by their action on lattice fields,
\begin{eqnarray}
(e^{aD_\tau}X)_\ell \;\equiv\; U_{\ell,\ell+1}\,X_{\ell+1}\,U_{\ell,\ell+1}^{-1},
\qquad
(e^{-aD_\tau}X)_\ell \;\equiv\; U_{\ell-1,\ell}^{-1}\,X_{\ell-1}\,U_{\ell-1,\ell}.
\label{eq:cov_shifts}
\end{eqnarray}
Equivalently, introducing the standard covariant forward/backward differences
\begin{eqnarray}
(D_+X)_\ell \;\equiv\;\frac{1}{a}\Big[(e^{aD_\tau}X)_\ell-X_\ell\Big]
=\frac{1}{a}\Big(U_{\ell,\ell+1}X_{\ell+1}U_{\ell,\ell+1}^{-1}-X_\ell\Big),
\label{eq:Dplus}\\
(D_-X)_\ell \;\equiv\;\frac{1}{a}\Big[X_\ell-(e^{-aD_\tau}X)_\ell\Big]
=\frac{1}{a}\Big(X_\ell-U_{\ell-1,\ell}^{-1}X_{\ell-1}U_{\ell-1,\ell}\Big),
\label{eq:Dminus}
\end{eqnarray}
one has \footnote{In \cite{Ydri2025} the opposite sign convention is used, so that
$\Delta = D_-D_+ \longrightarrow \partial_t^2$.}
\begin{eqnarray}
\Delta_{\Lambda,g} \;=-D_-D_+\longrightarrow -\partial_t^2 .
\label{eq:Delta_DmDp}
\end{eqnarray}
Hence
\begin{eqnarray}
a^2(\Delta_{\Lambda,g}X)_\ell
=
2X_\ell
-U_{\ell,\ell+1}X_{\ell+1}U_{\ell,\ell+1}^{-1}
-U_{\ell-1,\ell}^{-1}X_{\ell-1}U_{\ell-1,\ell}.
\label{eq:Delta_action}
\end{eqnarray}
Using cyclicity of the trace and relabelling lattice indices, one finds
\begin{eqnarray}
\frac{Na}{2}\sum_{\ell=0}^{\Lambda-1}\Tr\Big(X_\ell(\Delta_{\Lambda,g}X)_\ell\Big)
&=&\frac{N}{2a}
\sum_{\ell=0}^{\Lambda-1}
\Tr\Big(X_\ell-U_{\ell,\ell+1}X_{\ell+1}U_{\ell,\ell+1}^{-1}\Big)^2\nonumber\\
&=&\frac{N}{a}
\sum_{\ell=0}^{\Lambda-1}
\Tr\Big(X_\ell^2-X_\ell U_{\ell,\ell+1}X_{\ell+1}U_{\ell,\ell+1}^{-1}\Big).
\label{eq:XDeltX_square}
\end{eqnarray}
We now gauge--fix all links to unity except the closing link,
\begin{eqnarray}
U_{\ell,\ell+1}=\mathbf{1}\qquad(\ell=0,\dots,\Lambda-2),
\qquad
U_{\Lambda-1,0}=g,
\label{eq:gauge_fix_links}
\end{eqnarray}
so that \eqref{eq:XDeltX_square} becomes
\begin{eqnarray}
\frac{Na}{2}\sum_{\ell=0}^{\Lambda-1}\Tr\Big(X_\ell(\Delta_{\Lambda,g}X)_\ell\Big)
=
\frac{N}{a}\sum_{\ell=0}^{\Lambda-1}\Tr(X_\ell^2)
-\frac{N}{a}\sum_{\ell=0}^{\Lambda-2}\Tr(X_\ell X_{\ell+1})
-\frac{N}{a}\Tr\Big(X_{\Lambda-1}\,gX_0g^{-1}\Big).
\label{eq:XDeltX_holonomy_nosquare}\nonumber\\
\end{eqnarray}
Adding the mass term and restoring the overall normalization, the gauged lattice MHO action reads
\begin{eqnarray}
S_{\Lambda,g}[X]
&=&\frac{N}{2a}\sum_{\ell=0}^{\Lambda-1}\Tr X_\ell\Big((a^2\Delta_{\Lambda,g}+\frac{\beta^2}{\Lambda^2} s^2)X\Big)_\ell\nonumber\\
&=&\frac{N}{2a}\bigg[2\sum_{\ell=0}^{\Lambda-1}\Tr(X_\ell^2)
-2\sum_{\ell=0}^{\Lambda-2}\Tr(X_\ell X_{\ell+1})
-2\Tr\Big(X_{\Lambda-1}\,gX_0g^{-1}\Big)+\frac{\beta^2s^2}{\Lambda^2}\sum_{\ell=0}^{\Lambda-1}\Tr(X_\ell^2)\bigg].
\label{eq:lattice_MHO_holonomy}\nonumber\\
\end{eqnarray}
From \eqref{eq:lattice_MHO_holonomy} we read the $\Lambda N^2\times \Lambda N^2$ block tri--diagonal matrix
$M_{\Lambda,g}\equiv a^2\Delta_{\Lambda,g}+\frac{\beta^2}{\Lambda^2}s^2$ acting on the $\Lambda$-component vector
$X=(X_0,\ldots,X_{\Lambda-1})$.  In block form (each block is $N^2\times N^2$ acting by adjoint conjugation),
\begin{eqnarray}
M_{\Lambda,g}=
\left(
\begin{array}{cccccc}
(2+\mu^2)\,\mathbf{1} & -\mathbf{1} & 0 & \cdots & 0 & -\,{\rm Ad}_g^{-1}\\
-\mathbf{1} & (2+\mu^2)\,\mathbf{1} & -\mathbf{1} & \cdots & 0 & 0\\
0 & -\mathbf{1} & (2+\mu^2)\,\mathbf{1} & \ddots & \vdots & \vdots\\
\vdots & \ddots & \ddots & \ddots & -\mathbf{1} & 0\\
0 & \cdots & 0 & -\mathbf{1} & (2+\mu^2)\,\mathbf{1} & -\mathbf{1}\\
-\,{\rm Ad}_g & 0 & \cdots & 0 & -\mathbf{1} & (2+\mu^2)\,\mathbf{1}
\end{array}
\right),
\qquad
\mu^2\equiv\frac{\beta^2}{\Lambda^2}s^2.
\label{eq:M_matrix}\nonumber\\
\end{eqnarray}
Here $\mathbf{1}$ denotes the identity on the matrix space, while
\begin{eqnarray}
{\rm Ad}_g(X)\equiv gXg^{-1},
\qquad
{\rm Ad}_g^{-1}(X)\equiv g^{-1}Xg.
\end{eqnarray}
Gauge invariance and the normalization $\int d\mu(g)=1$ ensure that the path integral reduces to a single
integration over the holonomy $g$. The resulting partition function can therefore be written as
\begin{eqnarray}
Z_{N,\Lambda}
=
\int d\mu(g)\ {\bf Det}(M_{\Lambda,g})^{-1/2},
\label{eq:MolienWeyl_det}
\end{eqnarray}
where 
\begin{eqnarray}
M_{\Lambda,g}=a^2\Delta_{\Lambda,g}+\frac{\beta^2}{\Lambda^2} s^2=2-e^{aD_\tau}-e^{-aD_\tau}+\mu^2.
\end{eqnarray}

\medskip
\noindent The adjoint action can be represented as a tensor product. Indeed, the holonomy acts on the lattice fields in the adjoint representation,
\(
X \mapsto gXg^{-1},
\)
which defines a linear map \(\mathrm{Ad}_g:\mathrm{Mat}_N\to\mathrm{Mat}_N\).
Choosing the canonical matrix basis \(E_{ij}\) with
\((E_{ij})_{kl}=\delta_{ik}\delta_{jl}\), one finds
\begin{eqnarray}
\mathrm{Ad}_g(E_{ij})
&=&
gE_{ij}g^{-1}
=
\sum_{k,l} g_{ki}(g^{-1})_{jl}\,E_{kl}.
\end{eqnarray}
Upon vectorizing matrices according to
\(
\mathrm{vec}(E_{ij})=|i\rangle\otimes|j\rangle,
\)
this action is represented by
\begin{eqnarray}
\mathrm{vec}(gXg^{-1})
=
(g\otimes (g^{-1})^T)\,\mathrm{vec}(X).
\end{eqnarray}
Hence, in the vectorized adjoint space, the adjoint holonomy is represented by
the \(N^2\times N^2\) matrix
\begin{eqnarray}
\mathrm{Ad}_g \;=\; g\otimes (g^{-1})^T,
\end{eqnarray}
with inverse \(\mathrm{Ad}_g^{-1}=g^{-1}\otimes g\).

\medskip
\noindent 
Hence, in vectorized form of the lattice kernel
$M_{\Lambda,g}$ one may identify
${\rm Ad}_g=g\otimes(g^{-1})^T$ and
${\rm Ad}_g^{-1}=g^{-1}\otimes g^T$,
so that the upper--right corner is
$-(g^{-1}\otimes g^T)$ and the
lower--left corner is
$-(g\otimes(g^{-1})^T)$.
The diagonal blocks are $(2+\mu^2)\mathbf{1}$
and the nearest--neighbour off--diagonal blocks
are $-\mathbf{1}$.

\subsubsection{The $\Lambda\times\Lambda$ determinant and normal ordering}

\medskip
\noindent\textbf{The $\Lambda\times\Lambda$ determinant for $g=\mathbf{1}$.}

\medskip
\noindent For $g=\mathbf{1}$ the gauged lattice kernel is translationally invariant along the Euclidean
time lattice and takes the tri--diagonal nearest--neighbour form

\begin{eqnarray}
M_{\Lambda,1}
&=&
a^2\Delta_{\Lambda,1}+\frac{\beta^2}{\Lambda^2}s^2
=
(2+\mu^2)\,\mathbf{1}_\Lambda - T - T^{-1},
\label{eq:M_circulant}
\end{eqnarray}
where $T$ and $T^{-1}$ denote the forward and backward lattice translation operators,
$(TX)_\ell=X_{\ell+1}$ and $(T^{-1}X)_\ell=X_{\ell-1}$, with periodic identification
$X_{\ell\pm\Lambda}=X_\ell$. By translational invariance, $M_{\Lambda,1}$ is diagonalized by lattice momentum modes\footnote{Labeling lattice momenta by $k=1,\ldots,\Lambda$ is equivalent to the convention
$k=0,\ldots,\Lambda-1$ and avoids irrelevant phase shifts.}

\begin{eqnarray}
v^{(k)}_\ell=\exp\!\Big(\frac{2\pi i {k}}{\Lambda}\,\ell\Big),
\qquad
k=1,2,\ldots,\Lambda,
\end{eqnarray}
which satisfy
\begin{eqnarray}
T\,v^{(k)}_\ell = e^{\,\frac{2\pi i {k}}{\Lambda}}\,v^{(k)}_\ell,
\qquad
T^{-1}v^{(k)}_\ell = e^{-\,\frac{2\pi i {k}}{\Lambda}}\,v^{(k)}_\ell.
\end{eqnarray}
The corresponding eigenvalues of $M_{\Lambda,1}$ are therefore
\begin{eqnarray}
\lambda_k
&=&
(2+\mu^2)
- e^{\,\frac{2\pi i {k}}{\Lambda}}
- e^{-\,\frac{2\pi i {k}}{\Lambda}}
=
(2+\mu^2)
-2\cos\!\Big(\frac{2\pi {k}}{\Lambda}\Big),
\label{eq:eigs}
\end{eqnarray}
which can also be rewritten as
\begin{eqnarray}
\lambda_k
&=&
(2+\mu^2)-w_k-\frac{1}{w_k},\qquad w_k\equiv e^{\,\frac{2\pi i {k}}{\Lambda}}\nonumber\\
&=&\frac{1}{w_k}(z_+-w_k)(w_k-z_-),
\end{eqnarray}
where
\begin{eqnarray}
&&  z_+z_-=1\Leftrightarrow z_-=z_+^{-1}\nonumber\\
&&z_++z_-=2+\mu^2 \Leftrightarrow
z_\pm=\frac{1}{2}\Big((2+\mu^2)\pm\sqrt{(2+\mu^2)^2-4}\Big).
\label{eq:zpm_def}
\end{eqnarray}
The ``$\Lambda\times\Lambda$ lattice determinant'' then factorizes as
\begin{eqnarray}
{\rm Det}~ M_{\Lambda,1}
&=&
\prod_{k=1}^{\Lambda}
\Big[(2+\mu^2)-2\cos\!\Big(\frac{2\pi k}{\Lambda}\Big)\Big]\nonumber\\
&=&\prod_{k=1}^{\Lambda}(-w_k)^{-1}\prod_{k=0}^{\Lambda -1}(z_+-w_k)\prod_{k=0}^{\Lambda -1}(z_--w_k)\nonumber\\
&=&(-1)(z_+^{\Lambda}-1)(z_-^{\Lambda}-1)\nonumber\\
&=&z_+^{\Lambda}+z_-^{\Lambda}-2\nonumber\\
&=&z_+^{\Lambda}(1-z_-^{\Lambda})^2.\label{eq:det_product}
\end{eqnarray}

\medskip
\noindent{\bf Normal ordering and Molien--Weyl.}

\medskip
\noindent
The generalization of \eqref{eq:det_product} to $g\neq\mathbf{1}$ is straightforward in principle, though technically involved. It was shown in \cite{OConnor:2023mss} that the ``$\Lambda\times\Lambda$ lattice determinant'' of the kernel $M_{\Lambda,g}$ factorizes in general as

\begin{eqnarray}
{\rm Det}~ M_{\Lambda,g}
&=&
z_+^{\Lambda}
+ z_-^{\Lambda}
- g\otimes g^{-1}
- g^{-1}\otimes g
\nonumber\\
&=&
z_+^{\Lambda}
\big(1-z_-^{\Lambda}g\otimes g^{-1}\big)
\big(1-z_-^{\Lambda}g^{-1}\otimes g\big).
\label{eq:DetM}
\end{eqnarray}
The partition function of the gauged lattice theory is obtained by taking the remaining
``$N\times N$ matrix determinant'' over adjoint indices (denoted ${\bf det}$) and integrating over the gauge group,

\begin{eqnarray}
Z_{N,\Lambda}
&=&
\int \! \mu(g)\;
{\bf det}^{-1/2}~
\Big[(z_+^{\Lambda})(1-z_-^{\Lambda}g^{-1}\otimes g)
(1-z_-^{\Lambda}g\otimes g^{-1})\Big]
=
\int \! \mu(g)\;
\frac{z_-^{\frac{1}{2}N^2\Lambda}}
{{\bf det}~\big[1-z_-^{\Lambda}g\otimes g^{-1}\big]}.
\label{eq:ZN_Lambda}\nonumber\\
\end{eqnarray}
Here, we used the fact that the adjoint representation is
Hermitian,
\(
{\bf det}[1-z_-g\otimes g]
=
{\bf det}[1-z_-g^{-1}\otimes g].
\) Taking the continuum limit \(\Lambda\to\infty\) with \(\beta s\) fixed yields then 
\begin{eqnarray}
Z_{N,\infty}
&=&
\int \! \mu(g)\;
\frac{e^{-\frac{1}{2}N^2\beta s}}
{{\bf det}~\big[1-e^{-\beta s}g\otimes g^{-1}\big]}.
\label{eq:ZN_continuum}
\end{eqnarray}
The prefactor \(e^{-\frac{1}{2}N^2\beta s}\) arises from the zero--point energy of the harmonic
oscillators. Removing this factor corresponds to normal ordering and leads to the
Hilbert--Poincaré generating function in Molien--Weyl form,
\begin{eqnarray}
Z_{N,\infty}
&=&
\int \! \mu(g)\;
\frac{1}{{\bf det}~\big[1-x\,g\otimes g^{-1}\big]},\qquad x=e^{-\beta s}.
\label{eq:Molien_Weyl}
\end{eqnarray}

\medskip
\noindent
For the gauge group \(U(N)\), one may diagonalize the holonomy,
\(g=\mathrm{diag}(z_1,\dots,z_N)\) with \(z_i=e^{i\theta_i}\).
The Haar measure reduces to the Cartan torus with Vandermonde determinants
\begin{eqnarray}
\Delta(z)
=
\prod_{1\le i<j\le N}(z_i-z_j),
\qquad
\Delta(z^{-1})
=
\prod_{1\le i<j\le N}(z_i^{-1}-z_j^{-1}).\label{haar}
\end{eqnarray}
The Molien--Weyl formula becomes
\begin{eqnarray}
Z_{N,\infty}
=
\frac{1}{N!}
\oint
\prod_{i=1}^{N}\frac{dz_i}{2\pi i z_i}\;
\Delta(z)\Delta(z^{-1})
\prod_{i,j=1}^{N}
\frac{1}{1-x\,z_i z_j^{-1}}.
\label{eq:MW_explicit}
\end{eqnarray}

\subsection{Molien--Weyl formula for \(d>1\)}
\medskip
\noindent{\bf Normal--ordered $U(N)$ partition function.}

\medskip
\noindent Next, we consider a gauged matrix harmonic oscillator (MHO) with $d$ matrices $X_a$, defined by the action

\begin{eqnarray}
S_{\rm MHO}[X;\theta]
&=&
N\int_{0}^{\beta}\! dt\ \mathrm{Tr}\bigg[
\frac{1}{2}(D_tX_a)^2
+\frac{s^2}{2}X_a^2
\bigg],\label{MHO_action_for_Ward0}
\end{eqnarray}
Gauge fixing on the thermal circle reduces the gauge field $A_0$ to a constant holonomy

\begin{eqnarray}
g=\mathcal{P}\exp\Big(i\int_0^\beta dt\,A_0\Big)\in U(N),
\end{eqnarray}
so that the path integral becomes a single group integral over $g$. Since the $d$ matrices
factorize, the (normal--ordered) partition function is simply the $d$-th power of the
BFSS$_2$ Molien--Weyl integrand obtained in the previous section, viz.
\begin{eqnarray}
Z_{N,d}(x)
=
\int d\mu(g)\;
\frac{1}{\Big[{\bf det}\big(1-x\,g\otimes g^{-1}\big)\Big]^d},
\qquad x=e^{-\beta s}.
\label{eq:MW_BFSSd}
\end{eqnarray}
Diagonalizing $g=\mathrm{diag}(z_1,\ldots,z_N)$ gives the explicit Molien--Weyl form
\begin{eqnarray}
Z_{N,d}(x)
=
\frac{1}{N!}\oint\prod_{i=1}^{N}\frac{dz_i}{2\pi i z_i}\;
\Delta(z)\Delta(z^{-1})
\prod_{i,j=1}^{N}\frac{1}{\big(1-x\,z_i z_j^{-1}\big)^d},
\label{eq:MW_BFSSd_explicit}
\end{eqnarray}
which can equivalently be written as

\begin{eqnarray}
  Z_{N,d}(x)&=&\frac{1}{N!}\frac{1}{(1-x_b)^{n_bN}}\oint \prod_{i=1}^N\frac{dz_i}{2\pi i z_i}\Delta_A(-1,z)\frac{1}{\Delta_b^{n_b}(-x_b,z)}\nonumber\\
  &&x_b=e^{-\beta m_b}\equiv x, \qquad m_b\equiv s,\qquad n_b\equiv d.\label{eq:MW_BFSSd_explicit1}
\end{eqnarray}
The Faddeev--Popov--Vandermonde determinant $\Delta_A(1,z)$ and the bosonic determinant
$\Delta_B^{\,n_b}(x_b,z)$ are defined in terms of

 \begin{eqnarray}
   \Delta(x,z)=\prod_{i<j}(1+x\frac{z_i}{z_j})\prod_{i<j}(1+x\frac{z_j}{z_i}).
 \end{eqnarray}
 The factor  $1/|W|\equiv 1/N!$ in the partition function (\ref{eq:MW_BFSSd_explicit}) or (\ref{eq:MW_BFSSd_explicit1}) is part of the Haar measure normalization. It is the Weyl group volume, corresponding to division by permutations of the eigenvalues:
diagonal matrices related by permutations represent the same group element.

\medskip
\noindent{\bf $U(N)$ vs.\ $SU(N)$ in the Molien--Weyl formula.}

\medskip
\noindent It is clear that in the Molien--Weyl integral (\ref{eq:MW_BFSSd_explicit}), the terms with $i=j$ contribute an overall factor
\begin{equation}
\prod_{i=1}^{N}\frac{1}{(1-x_b)^d}=\frac{1}{(1-x_b)^{dN}},
\end{equation}
which corresponds to the $N$ zero--weights of the $U(N)$ adjoint.  Since
\(
\mathfrak{u}(N)=\mathfrak{su}(N)\oplus\mathfrak{u}(1),
\)
one of these zero--weights is associated with the decoupled $U(1)$ sector.
Projecting to the $SU(N)$ adjoint therefore amounts to removing a single
zero--weight factor, yielding
\begin{equation}
Z^{SU(N)}_{N,d}(x)
=
(1-x_b)^d\,Z^{U(N)}_{N,d}(x).
\end{equation}
We should then work with the formula
\begin{eqnarray}
  Z_{N,d}(x)&=&\frac{1}{N!}\frac{1}{(1-x_b)^{n_b(N-1)}}\oint \prod_{i=1}^N\frac{dz_i}{2\pi i z_i}\Delta_A(-1,z)\frac{1}{\Delta_b^{n_b}(-x_b,z)}.
\end{eqnarray}

\section{Molien--Weyl counting and BFSS$_2$--like factorization of singlet states}\label{section3}

\subsection{Explicit Molien--Weyl evaluation for $N=2$}

\subsubsection{Residue law}

\medskip
\noindent We now evaluate explicitly the Molien--Weyl integral for the bosonic BFSS$_{d+1}$ model at $N=2$.
This provides a nontrivial check of the holonomy formulation and clarifies the structure of the
$SU(2)$ singlet Hilbert space before proceeding to higher values of $N$.

For $U(2)$ we may diagonalize the holonomy as
\begin{eqnarray}
g=\mathrm{diag}(z_1,z_2), \qquad |z_i|=1 .
\end{eqnarray}
The Haar measure (\ref{haar}) for the group $U(2)$ reduces thus to a single contour integral with Vandermonde determinant
\begin{eqnarray}
\Delta(z)\Delta(z^{-1})=(1-z)(1-z^{-1}),\quad z=\frac{z_1}{z_2}.
\end{eqnarray}
For $d$ bosonic matrices of equal mass, the $U(2)$ Molien--Weyl partition function
\eqref{eq:MW_BFSSd_explicit} becomes then 
\begin{eqnarray}
Z^{\rm bos}_{2,d}(x)
&=&
\frac{1}{2}\oint_{|z|=1}\frac{dz}{2\pi i z}\;
(1-z)(1-z^{-1})
\prod_{i,j=1}^{2}\frac{1}{\big(1-x\,z_i z_j^{-1}\big)^d}.
\end{eqnarray}
Projecting to the $SU(2)$ adjoint amounts, as we have already point out earlier, to removing a single
zero--weight factor of $U(2)$, yielding the partition function 
\begin{eqnarray}
&&Z^{{\rm bos},SU(2)}_{2,d}(x)
\longrightarrow 
(1-x)^d\,Z^{{\rm bos},U(2)}_{2,d}(x)\nonumber\\
&&\Rightarrow Z^{\rm bos}_{2,d}(x)
=
\frac{(1-x)^d}{2}\oint_{|z|=1}\frac{dz}{2\pi i z}\;
(1-z)(1-z^{-1})
\prod_{i,j=1}^{2}\frac{1}{\big(1-x\,z_i z_j^{-1}\big)^d}.
\end{eqnarray}
Since $z_i z_j^{-1}\in\{1,z,z^{-1},1\}$, the product factorizes as
\begin{eqnarray}
\prod_{i,j=1}^{2}\frac{1}{\big(1-x\,z_i z_j^{-1}\big)^d}
=
\frac{1}{(1-x)^{2d}}\,
\frac{1}{(1-x z)^d(1-x z^{-1})^d}.
\end{eqnarray}
Hence the partition function may be written as
\begin{eqnarray}
Z^{\rm bos}_{2,d}(x)
&=&
\frac{1}{(1-x)^{d}}\;
\frac{1}{2}\oint_{|z|=1}\frac{dz}{2\pi i z}\;
\frac{(1-z)(1-z^{-1})}{(1-x z)^d(1-x z^{-1})^d}\equiv \frac{1}{(1-x)^{2d}}\;I_d(x).
\label{Z2d_MW_residue}
\end{eqnarray}
For $0<x<1$ the pole at $z=1/x$ lies outside the unit circle, while the factor
$(1-x z^{-1})^{-d}$ produces a pole at $z=x$ of order $d$ inside $|z|=1$.
In addition, after rewriting $(1-x z^{-1})^{-d}=(z/(z-x))^{d}$, one sees that
$z=0$ contributes only for $d=1$ (a simple pole), whereas for $d\ge 2$ the integrand
is regular at $z=0$.

We now evaluate,  using the residue theorem, the contour integral

\begin{eqnarray}
I_d(x)
=
-\frac{1}{2}\oint_{|z|=1}\frac{dz}{2\pi i}\;
\frac{z^{d-2}(z-1)^2}{(1-x z)^d\,(z-x)^d}.
\label{Id_simplified}
\end{eqnarray}

\subsubsection{Case $d=1$ (BFSS$_2$/BMN$_2$)}

\medskip
For $0<x<1$ the poles inside $|z|=1$ are at $z=0$ and $z=x$, both simple. From
\eqref{Id_simplified} with $d=1$,
\begin{eqnarray}
I_1(x)
=
-\frac{1}{2}\oint_{|z|=1}\frac{dz}{2\pi i}\;
\frac{(z-1)^2}{z(1-x z)(z-x)}
&=&\Res_{z=0}(\cdots)+\Res_{z=x}(\cdots)\nonumber\\
&=&-\frac{1}{2}\bigg[\frac{-1}{x}+\frac{1-x}{x(1+x)}\bigg]\nonumber\\
&=&\frac{1}{1+x}.
\end{eqnarray}
Hence
\begin{eqnarray}
Z^{\rm bos}_{2,1}(x)
=
\frac{1}{1-x}\,\frac{1}{1+x}
=
\frac{1}{1-x^2} \qquad {\rm BFSS}_2.
\label{Z21_result}
\end{eqnarray}

\subsubsection{General $d>1$ (BFSS$_{d+1}$/BMN$_{d+1}$): Cases $d=2$, $d=3$ and $d=5$}

\medskip
For $d>1$ the factor $z^{d-2}$ removes the would-be pole at $z=0$, and (for $0<x<1$) the only pole
inside the unit circle is $z=x$, now of order $d$. Using the standard residue formula for an
order-$d$ pole,
\begin{eqnarray}
I_d(x)
&=&
\Res_{z=x}\Bigg[
-\frac{1}{2}\,
\frac{z^{d-2}(z-1)^2}{(1-x z)^d\,(z-x)^d}
\Bigg]
\nonumber\\
&=&
-\frac{1}{2(d-1)!}\;
\left.
\frac{d^{\,d-1}}{dz^{\,d-1}}
\Bigg(
\frac{z^{d-2}(z-1)^2}{(1-x z)^d}
\Bigg)
\right|_{z=x}.
\label{Id_residue_general}
\end{eqnarray}
Therefore, for $d>1$,
\begin{eqnarray}
Z^{\rm bos}_{2,d}(x)
=
\frac{1}{(1-x)^{d}}\;
\Bigg[
-\frac{1}{2(d-1)!}\;
\left.
\frac{d^{\,d-1}}{dz^{\,d-1}}
\Bigg(
\frac{z^{d-2}(z-1)^2}{(1-x z)^d}
\Bigg)
\right|_{z=x}
\Bigg].\qquad {\rm BFSS}_{d+1}
\label{Z2d_residue_general}
\end{eqnarray}

\medskip
\noindent For $d=2$ and $d=3$ one finds from \eqref{Id_residue_general}
\begin{eqnarray}
I_2(x)&=&\frac{(1-x)^2}{(1-x^2)^3}\Rightarrow Z^{\rm bos}_{2,2}(x)
=\frac{1}{(1-x^2)^3}\qquad {\rm BFSS}_3.\label{bmn3}
\end{eqnarray}
\begin{eqnarray}
I_3(x)&=&\frac{(x-1)^2(x^2-x+1)}{(1-x^2)^5}\Rightarrow Z^{\rm bos}_{2,3}(x)
=\frac{x^2-x+1}{(1-x)(1-x^2)^5}=\frac{1+x^3}{(1-x^2)^6}\qquad {\rm BFSS}_4.\label{bmn4} \nonumber\\
\end{eqnarray}

\medskip
\noindent 
The calculation for \((N,d)=(2,5)\), corresponding to the \(N=2\) BFSS$_6$ model, is considerably more involved. 
The result, derived in the original arXiv version
of this work~\cite{Ydri:2026MolienV1}, is
 
\begin{eqnarray}
Z^{\rm bos}_{2,5}(x)
&=&
\frac{1}{(1-x^2)^{12}}\Big(
1+3x^2+10x^3+6x^4+6x^5+10x^6+3x^7+x^9
\Big).\label{bmn6}
\end{eqnarray}

\medskip
\noindent
Expanding the partition functions \eqref{Z21_result}, \eqref{bmn3}, \eqref{bmn4}, and
\eqref{bmn6} for \(d=1,2,3,5\) at small \(x\), one finds
\begin{eqnarray}
Z^{\rm bos}_{2,d}(x)
&=&
1+k\,x^2+\cdots,
\label{Z25_bos_smallx}
\end{eqnarray}
where the universal coefficient is
\begin{eqnarray}
k=\frac{d(d+1)}{2}=1,3,6,15.
\end{eqnarray}
This coefficient is equal to the number of independent gauge--invariant Gram operators
\(\Tr(X_aX_b)\), which dominate the very--low--temperature limit \(x\to0\).

In the BMN$_4$ and BMN$_6$ cases, which already display the generic behavior, this quadratic
term does \emph{not} arise from a simple global factor \((1-x^2)^{-k}\). Instead, it results from
nontrivial mixing between the Gram sector and additional singlet invariants. This mixing deforms
the naive BFSS$_2$--like factorized counting even in the uniform holonomy phase.

\subsection{Universal $x^2$ law and Gram--matrix dominance at very low $T$}

\subsubsection{Universal $x^2$ law at $N=2$}

\medskip
\noindent
We have obtained explicit closed forms for the bosonic \(N=2\) partition functions of several
BMN$_{d+1}$ models, i.e.\ mass--deformed BFSS$_{d+1}$ matrix quantum mechanical models, namely:

\begin{eqnarray}
Z^{\rm bos}_{2,1}(x)
=
\frac{1}{1-x^2} \qquad {\rm BFSS}_2.
\label{Z21_result1}
\end{eqnarray}

\begin{eqnarray}
Z^{\rm bos}_{2,2}(x)
=\frac{1}{(1-x^2)^3}\qquad {\rm BFSS}_3.\label{Z22_result1}
\end{eqnarray}
\begin{eqnarray}
Z^{\rm bos}_{2,3}(x)=\frac{1+x^3}{(1-x^2)^6}\qquad {\rm BFSS}_4.\label{Z23_result1}
\end{eqnarray}
\begin{eqnarray}
Z^{\rm bos}_{2,5}(x)=
\frac{1}{(1-x^2)^{12}}\Big(
1+3x^2+10x^3+6x^4+6x^5+10x^6+3x^7+x^9
\Big)\qquad {\rm BFSS}_6.\nonumber\\
\label{Z25_result1}
\end{eqnarray}
\medskip
\noindent In the very low--temperature regime ($x\to 0$), these partition functions admit a universal
expansion of the form

\begin{eqnarray}
Z^{\rm bos}_{2,d}(x)
&=&
1+\frac{d(d+1)}{2}x^2+\cdots,
\label{Z2d_bos_smallx}
\end{eqnarray}
independently of the detailed higher--order structure.

\medskip
\noindent This quadratic coefficient can be derived directly from the Molien--Weyl integral by
expanding the integrand to $O(x^2)$.

\medskip
\noindent 
Starting from the Molien--Weyl integrand,
\begin{eqnarray}
\prod_{i,j=1}^{N}\frac{1}{\big(1-x\,z_i z_j^{-1}\big)^{d}}
&=&
\exp\!\Bigg(
-d\sum_{i,j}\ln\big(1-x\,z_i z_j^{-1}\big)
\Bigg),
\end{eqnarray}
and using the expansion $-\ln(1-u)=\sum_{n\ge1}u^n/n$, one obtains
\begin{eqnarray}
-d\sum_{i,j}\ln\big(1-x\,z_i z_j^{-1}\big)
&=&
d\sum_{n\ge1}\frac{x^n}{n}\sum_{i,j}(z_i z_j^{-1})^n.
\end{eqnarray}
The double sum can be expressed in terms of group characters as
\begin{eqnarray}
\sum_{i,j}(z_i z_j^{-1})^n
&=&
\Tr(g^n)\,\Tr(g^{-n})
=
\chi_{\square}(g^n)\,\chi_{\overline{\square}}(g^n)
=
1+\chi_{\rm adj}(g^n),
\end{eqnarray}
where we used the $U(N)$ decomposition
\(\square\otimes\overline{\square}
=
\mathbf{1}\oplus \mathrm{Adj}_{U(N)}\).

\medskip
\noindent
The singlet contribution $\mathbf{1}$ is removed by the prefactor $(1-x)^d$, which projects
from $U(N)$ to $SU(N)$ by eliminating the center--of--mass sector.
Independently, the singlet component contained in the adjoint representation drops out
dynamically, since the adjoint action is implemented by commutators and the identity matrix
commutes trivially with all generators.

\medskip
\noindent
As a result, only the nontrivial $SU(N)$ adjoint degrees of freedom contribute to the
gauge--invariant spectrum. Consequently, when focusing on singlet operators built from
dynamical matrix degrees of freedom, one may equivalently interpret
$\chi_{\rm adj}$ as the adjoint character of $SU(N)$.

\medskip
\noindent 
Thus:
\begin{eqnarray}
(1-x)^d\prod_{i,j}\frac{1}{(1-x\,z_i z_j^{-1})^{d}}
=
\exp\!\left(
d\sum_{n\ge1}\frac{x^n}{n}\,\chi_{\rm adj}(g^n)
\right).
\label{plethystic_adj}
\end{eqnarray}
Expanding the exponent to second order in $x$,
\begin{eqnarray}
A(x)
&:=&
d\sum_{n\ge1}\frac{x^n}{n}\,\chi_{\rm adj}(g^n)
=
d\,x\,\chi_{\rm adj}(g)
+\frac{d}{2}x^2\chi_{\rm adj}(g^2)
+O(x^3),
\end{eqnarray}
and using $e^{A}=1+A+\frac12 A^2+\cdots$, one finds
\begin{eqnarray}
\prod_{i,j}\frac{1}{(1-x\,z_i z_j^{-1})^{d}}
=
1
+d\,x\,\chi_{\rm adj}(g)
+\frac{d}{2}x^2\chi_{\rm adj}(g^2)
+\frac{d^2}{2}x^2\chi_{\rm adj}(g)^2
+O(x^3).
\label{quadratic_character_terms}
\end{eqnarray}
Hence, we have 

\begin{eqnarray}
Z^{\rm bos}_{2,d}(x)
=
\int d\mu(g)\Bigg[
1
+d\,x\,\chi_{\rm adj}(g)
+\frac{d}{2}x^2\chi_{\rm adj}(g^2)
+\frac{d^2}{2}x^2\chi_{\rm adj}(g)^2
+O(x^3)
\Bigg].\label{quadratic_character_terms1}
\end{eqnarray}

\medskip
\noindent
For $SU(2)$, irreducible characters are orthonormal with respect to the normalized Haar measure,
\begin{eqnarray}
\int d\mu(g)\,\chi_j(g)\,\chi_{j'}(g) \;=\;\delta_{jj'},
\label{char_ortho}
\end{eqnarray}
where $j=0,\tfrac12,1,\ldots$ labels the spin of the irreducible representations of $SU(2)$.
Since the adjoint representation corresponds to spin $j=1$, while the singlet is $j=0$ with
$\chi_{j=0}(g)=1$, one immediately finds
\begin{eqnarray}
\int d\mu(g)\,\chi_{\rm adj}(g)
&=&
\int d\mu(g)\,\chi_{1}(g)\,\chi_{0}(g)=0,
\nonumber\\
\int d\mu(g)\,\chi_{\rm adj}(g)^2
&=&
\int d\mu(g)\,\chi_{1}(g)\,\chi_{1}(g)=1.
\nonumber
\end{eqnarray}
Moreover, since $g^2\in SU(2)$ and $\chi_{\rm adj}(g^2)$ is again the character of the spin--$1$
representation evaluated on a group element, orthogonality similarly implies
\begin{eqnarray}
\int d\mu(g)\,\chi_{\rm adj}(g^2)=1.
\end{eqnarray}

\medskip
\noindent
These results may also be derived explicitly by parametrizing an $SU(2)$ element in the fundamental as \(
g=\exp\!\big(i\theta\,\hat n\cdot\sigma\big),\)
\( \theta\in[0,\pi],\) \(\hat n\in S^2,\) so its eigenvalues are $e^{\pm i\theta}$ and therefore
\begin{eqnarray}
\chi_{\square}(g)=\Tr_{\square}(g)=e^{i\theta}+e^{-i\theta}=2\cos\theta.
\end{eqnarray}
The adjoint of $SU(2)$ is the spin--$1$ representation, obtained from
\begin{eqnarray}
\square\otimes\overline{\square}\;\cong\;\mathbf{1}\oplus {\rm adj},
\end{eqnarray}
For $SU(2)$ the fundamental representation is pseudoreal: the fundamental and anti-fundamental
representations are inequivalent as complex vector spaces, but they are equivalent as
group representations. Consequently, they have identical characters, and for the purpose of
character identities one may identify $\overline{\square}$ with $\square$. So, we have

\begin{eqnarray}
\chi_{\square}(g)\,\chi_{\overline{\square}}(g)
=
\chi_{\square}(g)^2
=
\chi_{\mathbf 1}(g)+\chi_{\rm adj}(g)
=
1+\chi_{\rm adj}(g).
\end{eqnarray}
Hence,  the spin--$1$ (adjoint) character is
\begin{eqnarray}
\chi_{\rm adj}(g)
=
\chi_{\square}(g)^2-1
=
(2\cos\theta)^2-1
=
1+2\cos(2\theta).
\label{chi_adj_SU2}
\end{eqnarray}
In this parametrization, the normalized Haar measure on $SU(2)$ is given by

\begin{eqnarray}
\int_{SU(2)} d\mu(g)\,f(g)
=
\frac{2}{\pi}\int_{0}^{\pi}d\theta\,\sin^2\theta\; f(\theta),
\qquad
\int_{SU(2)} d\mu(g)=1,
\label{Haar_SU2_theta}
\end{eqnarray}
where $f(\theta)$ denotes any class function (conjugation--invariant function), i.e.\ a function
that depends only on $\theta$ (equivalently on the eigenvalues $e^{\pm i\theta}$).

\medskip
\noindent 
We can then compute 

\begin{eqnarray}
\int d\mu(g)\,\chi_{\rm adj}(g)
=
\int d\mu(g)\,\big(1+2\cos(2\theta)\big)=
1+2\langle \cos(2\theta)\rangle
=
0.
\label{avg_chi1_zero}
\end{eqnarray}
\begin{eqnarray}
\int d\mu(g)\,\chi_{\rm adj}(g)^2
=
\int d\mu(g)\,\big(1+2\cos(2\theta)\big)^2&=&1+4\langle \cos(2\theta)\rangle+4\langle \cos^2(2\theta)\rangle\nonumber\\
&=&3+4\langle \cos(2\theta)\rangle+2\langle \cos(4\theta)\rangle=1.\nonumber\\
\label{avg_chi1_sq_one}
\end{eqnarray}

\medskip
\noindent
For the last identity, note that $g^2=\exp\!\big(i(2\theta)\hat n\cdot\sigma\big)$, hence
\begin{eqnarray}
\chi_{\rm adj}(g^2)=\chi_1(g^2)=1+2\cos(4\theta),
\label{chi1_g2}
\end{eqnarray}
and thus
\begin{eqnarray}
\int d\mu(g)\,\chi_{\rm adj}(g^2)
=\int d\mu(g)\,\big(1+2\cos(4\theta)\big)=1+2\langle \cos(4\theta)\rangle=1.
\label{avg_chi1_g2_one}
\end{eqnarray}
Using $SU(2)$ Haar orthogonality, one has therefore obtained 
\begin{eqnarray}
\int d\mu\,\chi_{\rm adj}(g)=0,
\qquad
\int d\mu\,\chi_{\rm adj}(g)^2=1,
\qquad
\int d\mu\,\chi_{\rm adj}(g^2)=1,
\end{eqnarray}
which reduces the $N=2$ bosonic partition function (\ref{quadratic_character_terms1}) to
\begin{eqnarray}
Z^{\rm bos}_{2,d}(x)
=
1+\left(\frac{d}{2}+\frac{d^2}{2}\right)x^2+O(x^3)
=
1+\frac{d(d+1)}{2}\,x^2+O(x^3).
\end{eqnarray}
Physically, the coefficient $d(d+1)/2$ counts the
independent quadratic adjoint singlets
$\Tr(X_a X_b)$, i.e.\ the Gram matrix operators.
For $N=2$, the adjoint representation of $SU(2)$
is the vector representation of $SO(3)$, and
these quadratic invariants are closely related
to the classical Gram construction of orthogonal
invariant theory
\cite{WeylClassicalGroups,ProcesiInvariantTheory}.
At very low temperature these operators dominate
the spectrum, even when additional higher--degree
invariants exist, as in BFSS$_4$ and BFSS$_6$.
This explains the universality of the quadratic
term and provides a robust benchmark for the
analysis at larger $d$ and in supersymmetric
extensions.

\subsubsection{Universal $x^2$ law at general $N$}

\medskip
\noindent
The universal quadratic counting law extends to
$SU(N)$ for every $N\geq 2$.
The relevant adjoint characters satisfy the
Haar identities
\begin{eqnarray}
\int d\mu(g)\,\chi_{\rm adj}(g) = 0,\qquad \int d\mu(g)\,\chi_{\rm adj}(g)^2 = 1,\qquad \int d\mu(g)\,\chi_{\rm adj}(g^2) = 1.
\end{eqnarray}
The last identity follows from
\begin{eqnarray}
\chi_{\rm adj}(g^2)
&=&
\Tr(g^2)\Tr(g^{-2})-1,
\nonumber\\
\int d\mu(g)\,
\Tr(g^2)\Tr(g^{-2})
&=&
2,
\qquad N\geq 2.
\end{eqnarray}
\medskip
\noindent
Expanding the bosonic Molien--Weyl integrand
to quadratic order in $x$ and applying these
identities yields
\begin{eqnarray}
Z^{\rm bos}_{N,d}(x)
&=&
1+\left(\frac{d}{2}+\frac{d^2}{2}\right)x^2
+O(x^3)
\nonumber\\
&=&
1+\frac{d(d+1)}{2}\,x^2+O(x^3),
\qquad N\geq 2.
\label{ZNd_quadratic_universal}
\end{eqnarray}
Thus, the leading singlet degeneracy is
universally determined by the
$d(d+1)/2$ independent quadratic Gram
operators $\Tr(X_aX_b)$, independently
of the gauge-group rank $N$.

\subsection{BFSS$_{d+1}$ singlets as BFSS$_2$--like towers (low $T$, $N=2$)}

\medskip
\noindent 
Throughout we work in the bosonic, normal--ordered Molien--Weyl setting with a single fugacity
\begin{eqnarray}
x=e^{-\beta s},
\end{eqnarray}
so that a monomial contributing $x^n$ corresponds to an excitation energy $E=n\,s$ in the effective
Gaussian description. For $N=2$, the holonomy integral projects onto gauge singlets and the spectrum
is read off from the small--$x$ expansion
\begin{eqnarray}
Z^{\rm bos}_{2,d}(x)=\sum_{n\ge 0} g_n^{(d)}\,x^n= \sum_{n\ge 0} g_n^{(d)}\, e^{-\beta E_n}
\qquad
g_n^{(d)}=\text{degeneracy at level }E_n=n\,s.
\end{eqnarray}

\subsubsection{$d=1$ (BFSS$_2$): only Gram invariants}

\medskip
\noindent 
With one matrix $X_1$, the only basic gauge--invariant building block is the quadratic ``Gram''
operator
\begin{eqnarray}
M_{11}=\Tr(X_1^2).
\end{eqnarray}
All singlets are powers of $M_{11}$, hence a single invariant tower. For $N=2$ one finds
\begin{eqnarray}
Z^{\rm bos}_{2,1}(x)=\frac{1}{1-x^2}
=1+x^2+x^4+\cdots,
\qquad
g_{2n}^{(1)}=1,\;\;g_{2n+1}^{(1)}=0,
\end{eqnarray}
i.e.\ equally spaced levels $E_{2n}=2ns$ with unit degeneracy.

\medskip
\noindent
Thus, the absence of odd powers of the fugacity $x$ in the $N=2$ bosonic partition function of
BFSS$_2$) (and in the $N=2$ partition function of BFSS$_3$) has a simple structural origin: at $N=2$ the gauge--invariant
singlet algebra at low orders is generated purely by \emph{quadratic} Gram invariants.

\subsubsection{$d=2$ (BFSS$_3$):  exact ``three--tower'' factorization at $N=2$}

\medskip
\noindent 
With two matrices $X_a$ ($a=1,2$), the quadratic Gram matrix
\begin{eqnarray}
M_{ab}:=\Tr(X_a X_b),\qquad a,b=1,2,
\end{eqnarray}
has $k=d(d+1)/2=3$ independent entries, namely $M_{11},M_{22},M_{12}$.
At quadratic order there is no additional $\epsilon$--invariant: since $M_{ab}$ is symmetric in
$(a,b)$ while $\epsilon_{ab}$ is antisymmetric, one has identically
\begin{eqnarray}
\epsilon_{ab}M_{ab}=0.
\end{eqnarray}
Moreover, there is no nontrivial cubic singlet for $d=2$. Any single--trace cubic is of the form
$\Tr(X_a X_b X_c)$, and with only two flavor indices $a,b,c\in\{1,2\}$ every contraction is
necessarily symmetric in at least two indices (by cyclicity of the trace), hence cannot be paired
with an antisymmetric $\epsilon_{ab}$. Equivalently, an $\epsilon$--type invariant first appears at
cubic order only when one has three matrices ($d\ge 3$), e.g.\ $\epsilon_{abc}\Tr(X_a X_b X_c)$.

\medskip
\noindent 
Thus, for $d=2$ the singlet spectrum at low orders is generated by the Gram invariants alone,
consistent with $Z^{\rm bos}_{2,2}(x)=(1-x^2)^{-3}$ at $N=2$.

\medskip
\noindent 
Concretely, for equal masses the $N=2$ bosonic answer is
\begin{eqnarray}
Z^{\rm bos}_{2,2}(x)=\frac{1}{(1-x^2)^3}
=\sum_{n\ge0}\binom{n+2}{2}\,x^{2n}.
\label{Z23_factorized}
\end{eqnarray}
This is \emph{exactly} the tensor product of three BFSS$_2$ Gram towers:
\begin{eqnarray}
\frac{1}{(1-x^2)^3}=\Big(\frac{1}{1-x^2}\Big)^3,
\end{eqnarray}
so the BFSS$_3$ spectrum is obtained by taking three independent copies of the BFSS$_2$ spectrum
and adding energies:
\begin{eqnarray}
E_{2n}=2ns,
\qquad
g^{(2)}_{2n}=\#\{(n_1,n_2,n_3)\in\mathbb{Z}_{\ge0}^3:\;n_1+n_2+n_3=n\}
=\binom{n+2}{2}.
\end{eqnarray}
This is the cleanest realization of the ``multiple BFSS$_2$--like invariant towers'' picture:
the three towers are the three independent Gram generators $(M_{11},M_{22},M_{12})$.

\subsubsection{$d=3$ (BFSS$_4$): $\epsilon_{abc}$ becomes independent and produces an $x^3$ channel}

\medskip
\noindent 
For three matrices $X_a$, $a=1,2,3$, the Gram matrix $M_{ab}$ has
\begin{eqnarray}
k=\frac{d(d+1)}{2}=6
\end{eqnarray}
independent quadratic invariants. In addition, there is now a genuinely new invariant with the
$SO(3)$ Levi--Civita tensor,
\begin{eqnarray}
{\cal E}_3 \;\sim\; \epsilon_{abc}\,\Tr(X_a X_b X_c),
\end{eqnarray}
which has no analogue at $d=1$ and  $d=2$.
This is precisely the mechanism that opens an \emph{odd} channel in the small--$x$ expansion: the
lowest $\epsilon$--type contribution appears at order $x^3$ (three letters).

\medskip
\noindent 
Nevertheless, the leading quadratic coefficient still matches the Gram counting:
the coefficient of $x^2$ is controlled by the $k=6$ quadratic Gram generators, so one still finds
\begin{eqnarray}
Z^{\rm bos}_{2,3}(x)=1+6x^2+\Big(\hbox{an }x^3\hbox{ term from }{\cal E}_3\Big)+\cdots,
\end{eqnarray}
i.e.\ the same $k=d(d+1)/2$ signature at the first nontrivial order, while the $\epsilon$--sector
first shows up at order $x^3$.

\subsubsection{\(d>3\) (BFSS$_6$, etc.)}

\medskip
\noindent
At very low temperature, \(x\to0\), the Molien--Weyl integral is dominated by the uniform
confining holonomy sector. In this regime, the first excited singlet level is also controlled by the
quadratic Gram data \(M_{ab}\), and therefore counts precisely the independent entries of the
Gram matrix:
\begin{eqnarray}
g^{(d)}_{2}=k=\frac{d(d+1)}{2}.
\end{eqnarray}
Higher invariants (including $\epsilon$--type ones) do exist for $d\ge 3$, but for the present
\(d>3\) cases they enter at higher orders in $x$ and/or through mixing with Gram composites. This is why, for example, in BFSS$_6$
($d=5$) one can find that the full rational answer is \emph{not} a pure $(1-x^2)^{-k}$ globally,
yet its small--$x$ expansion still begins as
\begin{eqnarray}
Z^{\rm bos}_{2,5}(x)=1+\frac{5\cdot 6}{2}\,x^2+\cdots=1+15x^2+\cdots,
\end{eqnarray}
because the Gram sector dominates the first nontrivial level even when additional invariant
families modify the detailed analytic structure at higher orders.

\medskip
\noindent The same residue/series technology can be pushed to higher $d$ (e.g.\ BFSS$_{10}$) and also to
supersymmetric models once the appropriate Molien--Weyl integrand (and its residue representation)
is fixed. In the SUSY case one expects additional fermionic towers and Clifford normalization
effects, but the Gram--matrix reorganization principle remains the guiding structure: BFSS$_{d+1}$
singlets reorganize into multiple BFSS$_2$--like towers, with the Gram sector controlling the
leading low--$T$ counting and $\epsilon$--type invariants appearing as genuinely new channels only
when $d$ is large enough to support them as independent operators.

\section{Supersymmetric extension}\label{section4}

\subsection{Supersymmetric Molien--Weyl integral at large $d$}

\medskip
\noindent We consider the large--mass limit of the BFSS$_{d+1}$/BMN$_{d+1}$ models. In this regime the quadratic
mass terms dominate, while the Yang--Mills interaction terms and, when present, the Chern--Simons
term are parametrically suppressed. On the fermionic side, the mass term similarly dominates over
the Yukawa interaction. The theory therefore reduces to a Gaussian gauged matrix
model with action of the generic form

\begin{eqnarray}
{S}_{{\rm BFSS}_{d+1}}= \int dt~ \mathrm{tr} \left( \frac{1}{2} D_t X^a D_t X_a - \frac{1}{2}\mu_b^2X_a^2  \right)-\frac{i}{2}\int dt~ \mathrm{tr} \bar\psi\bigg(\Gamma^t D_t +\mu_f\Gamma^F\bigg) \psi.\label{generic}
\end{eqnarray}
Here, $\mu_b$ and $\mu_f$ are the bosonic and fermionic masses, and $\Gamma^F$ is a fixed Clifford
matrix. Supersymmetry imposes a relation $M(\mu_b,\mu_f)=0$ between these masses. In the large--$d$
analysis, $\mu_b$ is replaced by the bosonic gap mass $s_b$, and consistency requires a corresponding
fermionic gap mass $s_f$ such that $M(s_b,s_f)=0$, yielding a supersymmetric completion of the large--$d$
bosonic saddle.

\medskip
\noindent The Gaussian model \eqref{generic} can be solved using the methods of
\cite{OConnor:2023mss,OConnor:2024udv}, leading to a normal--ordered $U(N)$ Molien--Weyl
representation

\begin{eqnarray}
Z_{\bf U(N)} = \int d\mu(g) \, \frac{{\bf det}\left(1 + x_f\, g \otimes g^{-1} \right)^{n_f}}{
{\bf det}\left(1 - x_{b1}\, g \otimes g^{-1} \right)^{n_{b1}} \, 
{\bf det}\left(1 - x_{b2}\, g \otimes g^{-1} \right)^{n_{b2}} }.\label{pf}
\end{eqnarray}
Here $d\mu(g)$ denotes the normalized Haar measure on $U(N)$. We have also allowed here for two bosonic species, with \(n_{b1}+n_{b2}=d,\) the total number of bosonic matrices. 
The fermions are encoded through $n_f$ independent
complex adjoint oscillator species. For each adjoint
component, the corresponding $2n_f$ Hermitian
Majorana combinations generate a Clifford algebra
of dimension $2^{2n_f}$, acting on a fermionic
Fock space of dimension $2^{n_f}$. The fugacities are related to the masses $m_{bi}$ and $m_f$ of the bosons and fermions by
\begin{eqnarray}
x_{b_i}=e^{-\beta m_{b_i}},\qquad x_f=e^{-\beta m_f}.
\end{eqnarray}

\medskip
\noindent
Projecting from $U(N)$ to $SU(N)$ removes the center--of--mass sector.
This is implemented by the standard prefactor
\[
\frac{(1-x_{b1})^{n_{b1}}(1-x_{b2})^{n_{b2}}}
{(1+x_f)^{n_f}},
\]
which yields the $SU(N)$ Molien--Weyl integral
\begin{eqnarray}
Z_{\bf SU(N)}
&=&
\frac{(1-x_{b1})^{n_{b1}}(1-x_{b2})^{n_{b2}}}
{(1+x_f)^{n_f}}
\int d\mu(g)\,
\frac{
{\bf det}\left(1+x_f\,g\otimes g^{-1}\right)^{n_f}
}{
{\bf det}\left(1-x_{b1}\,g\otimes g^{-1}\right)^{n_{b1}}
{\bf det}\left(1-x_{b2}\,g\otimes g^{-1}\right)^{n_{b2}}
}.\nonumber\\
\label{pf1}
\end{eqnarray}

\medskip
\noindent
The resulting Molien--Weyl integral may be written in contour form as
\begin{eqnarray}
Z_N^{(d)}(x_{b1},x_{b2},x_f)
&=&
\frac{1}{N!}
\frac{(1+x_f)^{n_f(N-1)}}
{(1-x_{b1})^{n_{b1}(N-1)}
 (1-x_{b2})^{n_{b2}(N-1)}}
\nonumber\\
&&\times
\oint\prod_{i=1}^{N}\frac{dz_i}{2\pi i z_i}\,
\Delta_{\rm A}(-1,z)
\frac{
\big[\Delta_{\rm f}(x_f,z)\big]^{n_f}
}{
\big[\Delta_{\rm b}(-x_{b1},z)\big]^{n_{b1}}
\big[\Delta_{\rm b}(-x_{b2},z)\big]^{n_{b2}}
}.
\end{eqnarray}
The Vandermonde determinant modified by an arbitrary parameter
$a$ is defined as
\begin{eqnarray}
\Delta(a,z)
=
\prod_{i<j}
\left(1+a\frac{z_i}{z_j}\right)
\left(1+a\frac{z_j}{z_i}\right).
\end{eqnarray}

\medskip
\noindent
The above expression is normal--ordered: the vacuum energy has
been subtracted. Restoring it amounts to multiplying by
\[
x_{b1}^{\frac{n_{b1}(N^2-1)}{2}}
x_{b2}^{\frac{n_{b2}(N^2-1)}{2}}
x_f^{-\frac{n_f(N^2-1)}{2}}.
\]
Thus,
\begin{eqnarray}
Z_N^{(d)}(x_{b1},x_{b2},x_f)
&\longrightarrow&
x_{b1}^{\frac{n_{b1}(N^2-1)}{2}}
x_{b2}^{\frac{n_{b2}(N^2-1)}{2}}
x_f^{-\frac{n_f(N^2-1)}{2}}
Z_N^{(d)}(x_{b1},x_{b2},x_f).
\end{eqnarray}

\subsection{$N=2$ supersymmetric Molien--Weyl residue formula}

\medskip
\noindent
For $U(2)$ we diagonalize
\begin{eqnarray}
g={\rm diag}(z_1,z_2),\qquad z_i=e^{i\theta_i},
\end{eqnarray}
so that in $g\otimes g^{-1}$ the eigenvalues are
\begin{eqnarray}
z_i z_j^{-1}\in\{1,1,z,z^{-1}\},\qquad z=\frac{z_1}{z_2}=e^{i(\theta_1-\theta_2)}.
\end{eqnarray}
\medskip
\noindent
Using the Weyl integration formula, i.e.\ the normalized $SU(2)$ Haar measure (the overall $U(1)$
drops since the integrand depends only on the ratio $z_1/z_2$), for any class function $f$ one has

\begin{eqnarray}
\int_{SU(2)} d\mu(g)\, f(g)
=
\frac{1}{2}\oint_{|z|=1}\frac{dz}{2\pi i z}\,
(1-z)(1-z^{-1})\, f(z),
\label{SU2_Weyl_z}
\end{eqnarray}
the $U(2)$ Gaussian SUSY Molien--Weyl integral becomes explicitly given by 
\begin{eqnarray}
Z^{(n_{b1},n_{b2};n_f)}_{U(2)}
=
\tfrac{1}{2}\oint_{|z|=1}\tfrac{dz}{2\pi i z}\,
(1-z)(1-z^{-1})\;
\tfrac{\Big[(1+x_f)^2(1+x_f z)(1+x_f z^{-1})\Big]^{n_f}}
{\Big[(1-x_{b1})^2(1-x_{b1} z)(1-x_{b1} z^{-1})\Big]^{n_{b1}}
 \Big[(1-x_{b2})^2(1-x_{b2} z)(1-x_{b2} z^{-1})\Big]^{n_{b2}}}.\nonumber\\
\label{Z_U2_z}
\end{eqnarray}

\medskip
\noindent
Projecting $U(2)\to SU(2)$ removes the extra $U(1)$ (center-of-mass) sector, i.e. it removes
\emph{one} of the two $(\cdots)^2$ singlet factors coming from the two unit eigenvalues in
$\{1,1,z,z^{-1}\}$. Equivalently,
\begin{eqnarray}
Z^{(n_{b1},n_{b2};n_f)}_{SU(2)}
=
\frac{(1-x_{b1})^{n_{b1}}(1-x_{b2})^{n_{b2}}}{(1+x_f)^{n_f}}\;
Z^{(n_{b1},n_{b2};n_f)}_{U(2)}.
\label{U2_to_SU2}
\end{eqnarray}
Substituting \eqref{Z_U2_z} into \eqref{U2_to_SU2} yields the clean $SU(2)$ contour form
\begin{eqnarray}
Z^{(n_{b1},n_{b2};n_f)}_{SU(2)}
&=&
\tfrac{1}{2}\tfrac{(1+x_f)^{n_f}}{(1-x_{b1})^{n_{b1}}(1-x_{b2})^{n_{b2}}}\oint_{|z|=1}\tfrac{dz}{2\pi i z}\,
(1-z)(1-z^{-1})\;
\tfrac{\Big[(1+x_f z)(1+x_f z^{-1})\Big]^{n_f}}
{\Big[(1-x_{b1} z)(1-x_{b1} z^{-1})\Big]^{n_{b1}}
  \Big[(1-x_{b2} z)(1-x_{b2} z^{-1})\Big]^{n_{b2}}}\nonumber\\
&=&\frac{(1+x_f)^{n_f}}{(1-x_{b1})^{n_{b1}}(1-x_{b2})^{n_{b2}}}I_{n_{b1},n_{b_2},n_f},\label{Z_SU2_z_final0}
\end{eqnarray}
where
\begin{eqnarray}
I_{n_{b1},n_{b_2},n_f}=-\frac{1}{2}\oint_{|z|=1}\frac{dz}{2\pi i}
\frac{z^{d-2-n_f}(z-1)^2\Big[(1+x_f z)(z+x_f)\Big]^{n_f}}
{\Big[(1-x_{b1} z)(z-x_{b1})\Big]^{n_{b1}}
  \Big[(1-x_{b2} z)(z-x_{b2})\Big]^{n_{b2}}}.
\label{Z_SU2_z_final}
\end{eqnarray}
For $|x_{b\alpha}|<1$ the poles inside $|z|=1$ are at
\begin{eqnarray}
z=0,\qquad z=x_{b1},\qquad z=x_{b2},
\end{eqnarray}
so \eqref{Z_SU2_z_final} is evaluated by the finite sum of residues at these points.

\subsection{Quadratic gauge singlets and universal quadratic law at low $T$}

\medskip
\noindent
The explicit residue evaluations of the $N=2$
supersymmetric Molien--Weyl integrals for
BFSS$_2$, BFSS$_3$, and BFSS$_4$ type-I
are presented in the original arXiv version
of this work~\cite{Ydri:2026MolienV1}.
The resulting closed expressions for the
$SU(2)$ singlet partition functions are:

\begin{eqnarray}
Z^{(1,0;1)}_{SU(2)}=
\frac{1+x_bx_f+x_bx_f^2+x_f^3}{1-x_b^2}\qquad {\bf BFSS}_2.
\end{eqnarray}

\begin{eqnarray}
Z^{(2,0;1)}_{SU(2)}
=
\frac{1 + x_f^3 + 2x_b x_f + 2x_b x_f^2 + x_b^2 x_f + x_b^2 x_f^2}
{(1-x_b^2)^3}\qquad {\bf BFSS}_3.
\end{eqnarray}

\begin{eqnarray}
Z^{(3,0;2)}_{SU(2)}(x_b,x_f)&=&\frac{(1+x_f)^2}{(1-x_b)(1-x_b^2)^5}\Big[
x_f^4-2x_f^3+4x_f^2-2x_f+1
-x_bx_f^4+8x_bx_f^3-7x_bx_f^2+8x_bx_f-x_b\nonumber\\
&+&x_b^2x_f^4-2x_b^2x_f^3+10x_b^2x_f^2-2x_b^2x_f+x_b^2
+2x_b^3x_f^2-2x_b^4x_f^2-x_b^5x_f^2\Big]\qquad {\bf BFSS}_4.\nonumber\\
\end{eqnarray}

\medskip
In the very low--temperature regime,
\begin{eqnarray}
x_b\ll1,\qquad x_f\ll1,
\end{eqnarray}
we expand the supersymmetric Molien--Weyl integrand to quadratic order in the two fugacities,
keeping the mixed term $x_bx_f$. For the three models computed above one immediately has:
\begin{eqnarray}
{\bf BFSS}_2:\quad d=1,\ n_f=1
&\Rightarrow&
Z=1+x_b^2+x_bx_f+O\!\Big((x_b,x_f)^3\Big),
\\
{\bf BFSS}_3:\quad d=2,\ n_f=1
&\Rightarrow&
Z=1+3x_b^2+2x_bx_f+O\!\Big((x_b,x_f)^3\Big),
\\
{\bf BFSS}_4\ {\rm type\ I}:\quad d=3,\ n_f=2
&\Rightarrow&
Z=1+6x_b^2+6x_bx_f+x_f^2+O\!\Big((x_b,x_f)^3\Big).
\end{eqnarray}
\medskip
These coefficients admit a direct operator interpretation. At quadratic order, gauge--invariant
singlets are adjoint bilinears built from Hermitian bosonic matrices $X_a$ ($a=1,\dots,d$) and
complex fermions $\psi_r$ ($r=1,\dots,n_f$). BFSS$_2$ is special: its single fermion is real
(Majorana), so it must be treated separately.

\medskip
\noindent Quadratic singlets fall into three classes:

\begin{itemize}

  \item \emph{The boson--boson bilinears:}  
The adjoint bosonic matrices $X_a$ ($a=1,\ldots,d$) are Hermitian. Gauge--invariant quadratic
singlets are therefore of the form $\Tr(X_a X_b)$. Since the trace is symmetric,
\begin{eqnarray}
\Tr(X_a X_b)=\Tr(X_b X_a),
\end{eqnarray}
only the symmetric combinations are independent. The number of such bilinears is
\begin{eqnarray}
\#\{\Tr(X_a X_b)\}=\frac{d(d+1)}{2},
\end{eqnarray}
which universally accounts for the coefficient of $x_b^2$ in the low--temperature expansion.

\item \emph{The boson--fermion bilinears:}  
For any BFSS model, each adjoint boson $X_a$ can be paired with each adjoint fermion $\psi_r$
to form a gauge--invariant bilinear $\Tr(X_a\psi_r)$. There is exactly one such invariant for
each choice of $(a,r)$, so their total number is
\begin{eqnarray}
\#\{\Tr(X_a\psi_r)\}=d\,n_f,
\end{eqnarray}
which directly explains the coefficient of the mixed term $x_bx_f$.

\item \emph{The fermion--fermion bilinears:}  
Here the distinction between real (Majorana) and complex fermions is crucial.

\begin{itemize}
\item \emph{BFSS$_2$:} There is a single real (Majorana) adjoint fermion, $n_f=1$.
Grassmann antisymmetry together with the reality condition implies
\begin{eqnarray}
\Tr(\psi\psi)=0,
\end{eqnarray}
so no fermion--fermion quadratic singlet exists and therefore no $x_f^2$ term appears.

\item \emph{BFSS$_3$:} Although the fermion can be viewed as complex, there is still only one
independent complex adjoint fermion ($n_f=1$). Grassmann antisymmetry again forbids
\begin{eqnarray}
\Tr(\psi\psi)=0,
\end{eqnarray}
so there is no $x_f^2$ contribution at quadratic order.

\item \emph{BFSS$_4$ (type I):} There are two independent complex adjoint fermions
($n_f=2$). In this case a single antisymmetric bilinear survives,
\begin{eqnarray}
\Tr(\psi_1\psi_2)=-\Tr(\psi_2\psi_1),
\end{eqnarray}
which yields exactly one independent fermion--fermion quadratic singlet and hence a single
$x_f^2$ term in the low--temperature expansion.
\end{itemize}

\end{itemize}

\medskip
\noindent Putting everything together, the very low--temperature expansion of any $N=2$
supersymmetric BFSS$_{d+1}$ model takes the universal form
\begin{eqnarray}
Z^{(d;n_f)}_{SU(2)}(x_b,x_f)
=
1
+\frac{d(d+1)}{2}\,x_b^2
+d\,n_f\,x_bx_f
+\frac{n_f(n_f-1)}{2}\,x_f^2
+O\!\Big((x_b,x_f)^3\Big),~d\geq 2,\label{Z_susy_universal_quadratic}\nonumber\\
\end{eqnarray}
independently of higher--order interactions and of the detailed structure of the full partition
function.  As in the purely bosonic case, the infrared spectrum is therefore dominated by
Gram--matrix--type bilinears and their supersymmetric extensions.

\subsection{Character derivation of the universal quadratic law}

\medskip
\noindent This universal quadratic structure admits a direct character derivation, parallel to the bosonic case.
Start from the $SU(2)$ integrand (after removing the center--of--mass singlet factors) in the
plethystic form:
\begin{eqnarray}
\frac{(1-x_b)^{d}}{(1+x_f)^{n_f}}\prod_{i,j}\frac{(1+x_f z_i z_j^{-1})^{n_f}}{(1-x_b z_i z_j^{-1})^{d}}
=
\exp\!\left(
A_b(x_b)+A_f(x_f)
\right),
\end{eqnarray}
with
\begin{eqnarray}
A_b(x_b)
&=&
d\sum_{n\ge1}\frac{x_b^n}{n}\,\chi_{\rm adj}(g^n),
\\
A_f(x_f)
&=&
n_f\sum_{n\ge1}\frac{(-1)^{n+1}}{n}\,x_f^n\,\chi_{\rm adj}(g^n).
\end{eqnarray}
Expanding to quadratic order gives
\begin{eqnarray}
A_b(x_b)
&=&
d\,x_b\,\chi_{\rm adj}(g)
+\frac{d}{2}\,x_b^2\,\chi_{\rm adj}(g^2)
+O(x_b^3),
\\
A_f(x_f)
&=&
n_f\,x_f\,\chi_{\rm adj}(g)
-\frac{n_f}{2}\,x_f^2\,\chi_{\rm adj}(g^2)
+O(x_f^3).
\end{eqnarray}
Using $e^{A_b+A_f}=1+(A_b+A_f)+\frac12(A_b+A_f)^2+\cdots$, the integrand becomes
\begin{eqnarray}
1
&+&
\Big(d\,x_b+n_f\,x_f\Big)\chi_{\rm adj}(g)
\nonumber\\
&+&
\frac{d}{2}x_b^2\,\chi_{\rm adj}(g^2)
-\frac{n_f}{2}x_f^2\,\chi_{\rm adj}(g^2)
+\frac{d^2}{2}x_b^2\,\chi_{\rm adj}(g)^2
+\frac{n_f^2}{2}x_f^2\,\chi_{\rm adj}(g)^2
+d\,n_f\,x_bx_f\,\chi_{\rm adj}(g)^2
+O\!\Big((x_b,x_f)^3\Big).\nonumber\\
\label{susy_integrand_quadratic}
\end{eqnarray}
Integrating over $SU(2)$ and using Haar orthogonality (adjoint is nontrivial)
\begin{eqnarray}
\int d\mu(g)\,\chi_{\rm adj}(g)=0,
\qquad
\int d\mu(g)\,\chi_{\rm adj}(g)^2=1,
\qquad
\int d\mu(g)\,\chi_{\rm adj}(g^2)=1,
\end{eqnarray}
one obtains immediately
\begin{eqnarray}
Z^{(d;n_f)}_{SU(2)}(x_b,x_f)
&=&
1
+\left(\frac{d}{2}+\frac{d^2}{2}\right)x_b^2
+\left(-\frac{n_f}{2}+\frac{n_f^2}{2}\right)x_f^2
+d\,n_f\,x_bx_f
+O\!\Big((x_b,x_f)^3\Big)
\nonumber\\
&=&
1
+\frac{d(d+1)}{2}\,x_b^2
+\frac{n_f(n_f-1)}{2}\,x_f^2
+d\,n_f\,x_bx_f
+O\!\Big((x_b,x_f)^3\Big),
\end{eqnarray}
which is \eqref{Z_susy_universal_quadratic}.

\medskip
\noindent In conclusion, the three quadratic contributions correspond to the three possible lowest singlet bilinears:
\begin{eqnarray}
\Tr(X_aX_b)\sim x_b^2,
\qquad
\Tr(X_a\psi_r)\sim x_bx_f,
\qquad
\Tr(\psi_r\psi_s)\sim x_f^2,
\end{eqnarray}
and the universal coefficients count these independent bilinear singlets:
\begin{eqnarray}
\#\{\Tr(X_aX_b)\}=\frac{d(d+1)}{2},
\qquad
\#\{\Tr(X_a\psi_r)\}=d\,n_f,
\qquad
\#\{\Tr(\psi_r\psi_s)\}=\frac{n_f(n_f-1)}{2}.\nonumber\\
\end{eqnarray}
Thus, at very low temperature the spectrum is dominated by bilinear invariants
(Gram--matrix operators and their supersymmetric extensions), irrespective of the presence of
higher--degree singlets.

\medskip
\noindent The character derivation presented above extends directly to $SU(N)$.  
Expanding the supersymmetric $U(N)$ Molien--Weyl integrand to quadratic order reproduces the
universal expansion \eqref{susy_integrand_quadratic}, with the understanding that the characters
are now those of $SU(N)$.  Upon Haar integration over $SU(N)$, the linear term vanishes since the
adjoint representation is nontrivial,

\begin{eqnarray}
\int d\mu(g)\,\chi_{\rm adj}(g)=0.
\end{eqnarray}
By character orthonormality one also has
\begin{eqnarray}
\int d\mu(g)\,\chi_{\rm adj}(g)^2=1.
\end{eqnarray}
It remains to evaluate the quadratic character $\chi_{\rm adj}(g^2)$.  Using
\begin{eqnarray}
\chi_{\rm adj}(g)=\Tr(g)\Tr(g^{-1})-1,
\qquad
\chi_{\rm adj}(g^2)=\Tr(g^2)\Tr(g^{-2})-1,
\end{eqnarray}
together with the standard second--moment identity for Haar $SU(N)$ (equivalently $U(N)$),
\begin{eqnarray}
\int d\mu(g)\,\Tr(g^2)\Tr(g^{-2})=2,
\qquad (N\ge2),
\end{eqnarray}
one finds
\begin{eqnarray}
\int d\mu(g)\,\chi_{\rm adj}(g^2)=1,
\qquad (N\ge2).
\end{eqnarray}
Substituting these Haar averages into \eqref{susy_integrand_quadratic} yields, for all $N\ge2$,
the universal quadratic law
\begin{eqnarray}
Z^{(d;n_f)}_{SU(N)}(x_b,x_f)
&=&
1
+\left(\frac{d}{2}+\frac{d^2}{2}\right)x_b^2
+\left(-\frac{n_f}{2}+\frac{n_f^2}{2}\right)x_f^2
+d\,n_f\,x_bx_f
+O\!\Big((x_b,x_f)^3\Big)
\nonumber\\
&=&
1
+\frac{d(d+1)}{2}\,x_b^2
+\frac{n_f(n_f-1)}{2}\,x_f^2
+d\,n_f\,x_bx_f
+O\!\Big((x_b,x_f)^3\Big),
\end{eqnarray}
Thus the quadratic coefficients
are independent of $N$ and depend only on $(d,n_f)$.  Equivalently, at very low temperature the
spectrum is universally dominated by bilinear gauge--invariant singlets (Gram--matrix operators
and their supersymmetric extensions), irrespective of the presence of higher--degree invariants
at finite temperature.

\section{Hamiltonian derivation of the \(N=2\) exact BFSS$_2$--factorization of BFSS$_3$}\label{section5}

\subsection{Exact BMN$_2$ decomposition of BMN$_{d+1}$ at $(d,N)=(2,2)$}
\medskip
\noindent
We consider BFSS$_{d=1}$ matrix quantum mechanics with maximal supersymmetric
mass deformation, i.e.\ its BMN$_{d+1}$ extension.
Throughout this section we focus on the special point $(d,N)=(2,2)$, where an
exact factorization property holds.

\medskip
\noindent
Although the fermion in BFSS$_2$ is \emph{real} (Majorana), the correct quantity
entering the partition function is still $n_f=1$, exactly as in BFSS$_3$.
The reason is that $n_f$ counts \emph{fermionic oscillators}, i.e.\ creation--annihilation
pairs $(b,b^\dagger)$ appearing in the Hamiltonian, rather than the number of real
spinor components.
This is completely analogous to the bosonic counting, where
$d=n_b$ counts \emph{bosonic oscillators}, i.e.\ the number of pairs
$(a_a,a_a^\dagger)$.

\medskip
\noindent
From the Hamiltonian point of view, BFSS$_2$ and BFSS$_3$ are therefore constructed
on the same fermionic Fock space: in both cases quantization produces a single
fermionic oscillator $(b,b^\dagger)$.
The only difference between the two models lies in the mass assignment:
$m_f=0$ in the mass-deformed BFSS$_2$ (as required by supersymmetry in $d=1$), whereas $m_f\neq0$
in the mass-deformed BFSS$_3$.
No difference arises at the level of oscillator counting.

\medskip
\noindent
More precisely, for BFSS$_2$ the fermion is massless, $m_f=0$, and
the real--fermion realization is obtained by retaining
a single $Q_+$--invariant singlet tower of the auxiliary
complex Fock space, as discussed below. For BFSS$_3$
the fermion is massive, $m_f\neq 0$, and the full complex
fermionic Fock space is retained. Thus the initial
oscillator counting is the same in both constructions.

\medskip
\noindent
In a previous section we evaluated the $N=2$ partition functions of the mass-deformed BFSS$_2$ and
BFSS$_3$ using the Molien--Weyl bosonic integrals \eqref{Z2d_MW_residue} and \eqref{Id_simplified} and the Molien--Weyl supersymmetric integrals
\eqref{Z_SU2_z_final0} and \eqref{Z_SU2_z_final}.
These results agree with those obtained previously by Denjoe~O'Connor
using \emph{Mathematica} \cite{OConnor_private}.
All partition functions quoted below are normal--ordered, and we use the standard fugacities
\begin{eqnarray}
x_b:=e^{-\beta m_b},
\qquad
x_f:=e^{-\beta m_f}.
\end{eqnarray}
For BFSS$_3$ one has $m_b=\mu/6$, $m_f=\mu/4$, while for BFSS$_2$ one sets
$m_b=\mu/6=\sqrt{-\Lambda}$ and (for a real fermion) $m_f=0$.

\medskip
\noindent
The basic result for BFSS$_2$ is the $SU(2)$ singlet partition function with one
bosonic letter of fugacity $x_b$ and one fermionic letter of fugacity $x_f$,
\begin{eqnarray}
Z_{2,2}(x_b,x_f)
:=
\frac{1 + x_b x_f + x_b x_f^2 + x_f^3}{1 - x_b^2}.
\label{Denjoe_SfAbSU2D1}
\end{eqnarray}
Setting $x_f=0$ yields the purely bosonic partition function
\begin{eqnarray}
Z_{2,2}(x,0)=\frac{1}{1-x^2},
\label{BFSS2_bos_N2}
\end{eqnarray}
while in the supersymmetric case $m_f=0$ so that $x_f=1$, giving
\begin{eqnarray}
Z_{2,2}(x,1)
=
\frac{2}{1-x}.
\end{eqnarray}
Dividing by $2^{N-1}$, the dimension of the Clifford algebra of the $N-1$ real
fermionic zero modes, normalizes the zero--temperature partition function to unity,
and one obtains
\begin{eqnarray}
Z_{2,2}^{\rm susy}(x)
=
\frac{1}{1-x}.
\label{BFSS2_susy}
\end{eqnarray}

\medskip
\noindent
The corresponding $SU(2)$ singlet partition function for BFSS$_3$
(two bosonic letters with fugacities $x_{b_1},x_{b_2}$ and one fermionic letter
with fugacity $x_f$) is
\begin{eqnarray}
Z_{2,3}
=
\frac{(1+x_f)\,
\big(1-x_f+x_{b_1}x_f+x_{b_2}x_f+x_{b_1}x_{b_2}x_f+x_f^2\big)}
{(1-x_{b_1}^2)(1-x_{b_2}^2)(1-x_{b_1}x_{b_2})}.
\label{Denjoe_SfSU2D2}
\end{eqnarray}
Setting  $x_{b_1}=x_{b_2}=x$ gives the supersymmetric result, while setting $x_f=0$ and $x_{b_1}=x_{b_2}=x$ gives the purely bosonic result
\begin{eqnarray}
Z_{2,3}(x,0)
=
\frac{1}{(1-x^2)^3}.
\label{BFSS3_bos_general}
\end{eqnarray}

\medskip
\noindent
\textbf{Theorem (Exact factorization at $(d,N)=(2,2)$).}
For the bosonic theory at $N=2$ one has the exact identity
\begin{eqnarray}
Z_{2,3}^{\rm bos}
=
\big(Z_{2,2}^{\rm bos}\big)^3.
\label{Exact_factorization}
\end{eqnarray}
This factorization is exact, holds at all temperatures, and is special to
$(d,N)=(2,2)$. Nevertheless, the bosonic Gaussian BFSS$_3$ model at $N=2$
exhibits a smooth deconfinement--like holonomy rearrangement, without
a genuine phase transition.
This is consistent with generic finite--$N$ models, where the
large--$N$ phase transition typically persists as a smooth crossover
away from the $N\to\infty$ limit \cite{Ydri2025}.
In fact, the special point $(d,N)=(2,2)$ shares the exact factorized
singlet thermodynamics of three BFSS$_2$ copies, but this does not
preclude a qualitative rearrangement of the holonomy distribution
with temperature.

\medskip
\noindent
In the remainder of this section we give an independent Hamiltonian derivation
of \eqref{Exact_factorization} by explicitly constructing the $SU(2)$ singlet
Hilbert space and enforcing the Gauss law
\begin{eqnarray}
{\cal J}_A|{\rm phys}\rangle=0.
\end{eqnarray}
This provides a direct Hilbert--space interpretation of the Molien--Weyl
result and clarifies why the exact factorization is a special feature of
$(d,N)=(2,2)$ and does not persist for higher $d$ and $N$.

\medskip
\noindent
We next clarify the distinction between Hagedorn and deconfinement phase
transitions, and emphasize the crucial difference between the
infinite--$N$ limit and fixed--$N$ models. We show that only in the
$N\rightarrow\infty$ limit can competing holonomy saddles produce a genuine
nonanalyticity in the free energy. At any fixed finite $N$, the partition
function remains analytic for all temperatures, and the large--$N$
transition can survive at most as a smooth crossover.

\medskip
\noindent
The special case $(d,N)=(2,2)$ is distinguished
by exact BFSS$_2$ singlet factorization, but
this does not exclude a smooth
deconfinement--like holonomy crossover.

\subsection{Hilbert--space derivation of $N=2$ BFSS$_2$/BMN$_2$}
\subsubsection{Setup}
\medskip
\noindent
We will rederive here the $SU(2)$ partition functions of BFSS$_2$ and BFSS$_3$ by
counting directly the \emph{gauge--singlet} states in the oscillator Fock space,
i.e.\ by enforcing the underlying Gauss law ${\cal J}_A|{\rm phys}\rangle=0$.

\medskip
\noindent
For $SU(2)$ the adjoint representation is the real three--vector representation.
Thus each bosonic matrix degree of freedom is equivalently a triplet of oscillators
$a_{aA}^{\dagger}$, where $A=1,2,3$ is rotated by the gauge group and
$a=1,\ldots,d$ labels the matrix index.
Gauge--invariant operators are obtained by contracting adjoint indices with
$\delta_{AB}$ (and, when available, $\epsilon_{ABC}$).

\medskip
\noindent
The fermionic operators $b_A,b_A^\dagger$ denote complex
adjoint oscillators satisfying
\begin{equation}
\{b_A,b_B^\dagger\}=\delta_{AB},
\qquad
\{b_A,b_B\}=\{b_A^\dagger,b_B^\dagger\}=0.
\end{equation}
They are used to construct the fermionic Fock space and
its decomposition into exterior powers of the adjoint
representation.

\medskip
\noindent
For bosons $a_{aA}$ and fermions $b_A$ in the adjoint ($A=1,2,3$) we take the following standard Hamiltonian

\begin{eqnarray}
H=\frac{1}{2}{\rm Tr}\bigg[m_b\sum_{a=1}^da_a^{\dagger}a_a+m_fb^{\dagger}b\bigg]=
m_b\, \sum_{a=1}^da_{aA}^{\dagger}a_{aA}
+
m_f\, b_A^{\dagger}b_A.
\end{eqnarray}
The partition function is defined by
\begin{eqnarray}
Z \;=\; {\rm Tr}_{\rm phys}\, e^{-\beta H}.
\end{eqnarray}
The Gauss constraint
  \begin{eqnarray}
[P,X]+i\psi\psi\equiv 0
    \end{eqnarray}
embodies the fact that physical states are $SU(2)$ singlets.
In oscillator language, the corresponding adjoint gauge generators are

\begin{eqnarray}
{\cal J}_A
=
-i\epsilon_{ABC}
\Big(
\sum_{a=1}^d a_{aB}^{\dagger}a_{aC}
+
b_B^{\dagger}b_C
\Big),
\qquad
{\cal J}_A|{\rm phys}\rangle=0.
\label{Gauss_SU2}
\end{eqnarray}
\subsubsection{Single-matrix bosonic Hamiltonian}
\medskip
\noindent
For a single bosonic matrix ($d=1$), the creation operators $a_A^{\dagger}$
transform in the adjoint (spin--$1$) irrep $V_1$ of dimension $3$. A state with fixed boson number $n$ is then obtained by acting with $n$
bosonic creation operators on the vacuum, and therefore transforms in
the completely symmetric tensor product
\begin{eqnarray}
{\cal H}^{(n)}_{\rm bos}\;\simeq\;{\rm Sym}^n(V_1).
\end{eqnarray}
Gauge invariance requires physical states to be $SU(2)$ singlets,
so we must extract the trivial representation from ${\rm Sym}^n(V_1)$. A standard Clebsch--Gordan analysis shows that the singlet appears
with multiplicity
\begin{eqnarray}
{\rm mult}\big(\mathbf{1}\ \text{in}\ {\rm Sym}^n(V_1)\big)
=
\left\{
\begin{array}{ll}
1, & n\ \text{even},\\[1mm]
0, & n\ \text{odd}.
\end{array}
\right.
\label{singlet_SymnV1}
\end{eqnarray}
Hence, gauge--invariant bosonic states exist only at even occupation
number $n=2k$.

\medskip
\noindent
Rotational invariance then fixes the structure of the singlet uniquely:
for $n=2k$ the only possible invariant is obtained by contracting the
adjoint indices pairwise with $\delta_{AB}$. Equivalently, the unique
singlet state at level $n=2k$ is generated by repeated action of the
quadratic invariant
\begin{eqnarray}
K_+ \;:=\; a_A^{\dagger}a_A^{\dagger},
\end{eqnarray}
so that the physical singlet states are
\begin{eqnarray}
(K_+)^k|0\rangle,
\qquad k=0,1,2,\ldots
\qquad (n=2k).
\end{eqnarray}
Using \eqref{singlet_SymnV1}, we obtain the $N=2$ bosonic partition function of BFSS$_2$ as follows 
\begin{eqnarray}
Z^{\rm bos}_{2}=
{\rm Tr}_{\rm phys}\, x_b^{\hat N}
&=&
\sum_{n\geq 0}\Big[{\rm mult}(\mathbf{1}\ {\rm in}\ {\rm Sym}^n(V_1))\Big]\,x_b^{n}
\nonumber\\
&=&\sum_{k\geq 0} x_b^{2k}=1+x_b^2+x_b^4+...\nonumber\\
&=&
\frac{1}{1-x_b^2}.
\label{Z2_bos_SU2}
\end{eqnarray}

\medskip
\noindent
In summary, we have:
\begin{itemize}
\item The first term in  the series corresponds to the vacuum $|0\rangle$.

\item  The second term in the series corresponds to the state $a_C^{\dagger}a_C^{\dagger}|0\rangle=K_+|0\rangle$. The state $a_C^{\dagger}|0\rangle$ is not allowed by the Gauss constraint.

\item  The third term corresponds to the state $(a_C^{\dagger}a_C^{\dagger})^2|0\rangle=K_+^2|0\rangle$.
\end{itemize}
The first few physical bosonic singlets are therefore 
\begin{eqnarray}
|0\rangle,\qquad
K_+|0\rangle,\qquad
K_+^2|0\rangle,\ \ldots
\end{eqnarray}
and there is \emph{no} one-boson singlet, since $V_1$ contains no trivial representation.

\subsubsection{Single-matrix supersymmetric Hamiltonian}

\medskip
\noindent We consider now the addition of one complex adjoint fermion.
The fermionic creation operators $b_A^{\dagger}$ transform in the adjoint
(spin--$1$) representation $V_1$ of $SU(2)$.
At fixed fermion number $k$, the fermionic Fock space transforms as the
exterior power $\Lambda^k(V_1)$,
\begin{eqnarray}
\Lambda^0(V_1)=\mathbf{1},\qquad
\Lambda^1(V_1)=V_1,\qquad
\Lambda^2(V_1)=V_1,\qquad
\Lambda^3(V_1)=\mathbf{1},
\label{ext_powers_V1}
\end{eqnarray}
with $\Lambda^k(V_1)=0$ for $k\ge4$ due to Grassmann antisymmetry.

\medskip
\noindent
In more detail, we have:
\begin{itemize}
  \item \textbf{Explicit realization of $\Lambda^0(V_1)\simeq \mathbf{1}$.} 
By definition, $\Lambda^0(V_1)$ is the one--dimensional trivial representation.
In the fermionic Fock space it is spanned by the vacuum state
\(|0\rangle,\)
which is annihilated by all Gauss generators,
\({\cal J}_A|0\rangle=0.\)
Hence the zero--fermion sector transforms as the singlet representation
$\mathbf{1}$ of $SU(2)$.

\item \textbf{Explicit realization of $\Lambda^1(V_1)\simeq V_1$.} 
In the fermionic Fock space, the one--fermion sector is spanned by
\begin{eqnarray}
|A\rangle \;:=\; b_A^{\dagger}|0\rangle,\qquad A=1,2,3,
\end{eqnarray}
and the Gauss generators act by the adjoint (vector) action
\begin{eqnarray}
[{\cal J}_D,b_A^{\dagger}]
=
i\,\epsilon_{DAE}\,b_E^{\dagger}
\;\;\Rightarrow\;\;
{\cal J}_D|A\rangle
=
i\,\epsilon_{DAE}\,|E\rangle.
\end{eqnarray}
Hence the $k=1$ fermion subspace carries exactly the $V_1$ irrep.

  \item \textbf{Explicit realization of $\Lambda^2(V_1)\simeq V_1$.}
For $SU(2)$ the adjoint representation $V_1$ is the real three--vector
representation, equipped with the invariant tensors $\delta_{AB}$ and
$\epsilon_{ABC}$. As a consequence, the exterior square $\Lambda^2(V_1)$
is equivalent to $V_1$ as an $SU(2)$ representation.
This equivalence can be realized explicitly using the $\epsilon$--tensor.
Given two adjoint fermionic creation operators, the antisymmetric
two--fermion state \(b_B^{\dagger}b_C^{\dagger}|0\rangle\) may be dualized to a single adjoint index as follows:
\begin{eqnarray}
b_B^{\dagger}b_C^{\dagger}|0\rangle\longrightarrow \tilde b_A^{\dagger}
\;:=\;
\frac{1}{2}\,\epsilon_{ABC}\,b_B^{\dagger}b_C^{\dagger}.
\label{btilde_def}
\end{eqnarray}
The operators $\tilde b_A^{\dagger}$ transform as a vector under
$SU(2)$, viz. \([{\cal J}_D,\tilde b_A^{\dagger}]=i\,\epsilon_{DAE}\,\tilde b_E^{\dagger},\)
and the map \eqref{btilde_def} is invertible up to normalization.
Hence $\Lambda^2(V_1)$ and $V_1$ are isomorphic as $SU(2)$ representations,
not merely as vector spaces.

\item \textbf{Explicit realization of $\Lambda^3(V_1)\simeq \mathbf{1}$.}
The top exterior power of the adjoint $SU(2)$ representation $V_1\simeq\mathbb{R}^3$
is one--dimensional and transforms in the determinant representation.
Since $SU(2)$ acts through $SO(3)$ on $V_1$, the determinant is $+1$, hence
$\Lambda^3(V_1)$ is the trivial irrep $\mathbf{1}$.
Concretely, the unique (up to normalization) three--fermion state is obtained by
contracting with the invariant Levi--Civita tensor:
\begin{eqnarray}
|{\bf 1}\rangle_{(3)}
\;:=\;
\frac{1}{3!}\,\epsilon_{ABC}\,b_A^{\dagger}b_B^{\dagger}b_C^{\dagger}|0\rangle.
\label{Lambda3_scalar_state}
\end{eqnarray}
This is a gauge singlet, namely \({\cal J}_D\,|{\bf 1}\rangle_{(3)}=0\), and thus $\epsilon_{ABC}b_A^{\dagger}b_B^{\dagger}b_C^{\dagger}|0\rangle$
spans $\Lambda^3(V_1)$ and is an explicit scalar (singlet) representative.

\end{itemize}

\medskip
\noindent
To construct physical states, we couple the fermionic representation
$\Lambda^k(V_1)$ to the bosonic sector and impose the singlet condition.
This yields one basic singlet representative for each fermion number $k$:
\begin{itemize}
\item $k=0$: The fermionic vacuum is already a singlet and
contributes $1$.

\item $k=1$: The fermionic sector transforms as $V_1$ and
can be coupled to one bosonic $V_1$ to form a singlet.
Since $V_1\otimes V_1$ contains the trivial representation
once, this contributes $x_bx_f$. A representative is
$a_A^\dagger b_A^\dagger|0\rangle$.

\item $k=2$: Since $\Lambda^2(V_1)\simeq V_1$, the
two--fermion sector similarly combines with one bosonic
adjoint to form a singlet, contributing $x_bx_f^2$.
A representative is
$\epsilon_{ABC}a_A^\dagger b_B^\dagger b_C^\dagger|0\rangle$.

\item $k=3$: The top exterior power
$\Lambda^3(V_1)\simeq\mathbf{1}$ is already a singlet
and contributes $x_f^3$.
\end{itemize}

\medskip
\noindent
Multiplying by the bosonic singlet tower
\eqref{Z2_bos_SU2}, we obtain the supersymmetric
$SU(2)$ partition function
\begin{equation}
Z_{SU(2)}^{\rm susy}
=
{\rm Tr}_{\rm phys}\,
x_b^{\hat N_b}x_f^{\hat N_f}
=
\frac{1+x_bx_f+x_bx_f^2+x_f^3}
{1-x_b^2}.
\label{Z_SU2_complexfermion}
\end{equation}

\medskip
\noindent
A convenient choice of basic singlet representatives is
\begin{eqnarray}
k=0:&\quad& |0\rangle,\nonumber\\
k=1:&\quad& a_A^\dagger b_A^\dagger|0\rangle,\nonumber\\
k=2:&\quad&
\epsilon_{ABC}a_A^\dagger b_B^\dagger
b_C^\dagger|0\rangle,\nonumber\\
k=3:&\quad&
\epsilon_{ABC}b_A^\dagger b_B^\dagger
b_C^\dagger|0\rangle.
\end{eqnarray}
Each representative can be dressed by arbitrary powers
of the bosonic singlet operator
\begin{equation}
K_+:=a_A^\dagger a_A^\dagger,
\end{equation}
giving four independent gauge--singlet towers.

\medskip
\noindent
To connect this complex--fermion Fock construction with
the Majorana Clifford realization, introduce
\begin{equation}
\gamma_A:=b_A^\dagger+b_A,
\qquad
\{\gamma_A,\gamma_B\}=2\delta_{AB}.
\end{equation}
The Clifford realization of the raising supercharge is
\begin{equation}
Q_+:=a_A^\dagger\gamma_A.
\end{equation}
Since $b_A|0\rangle=0$, its action on the vacuum gives
\begin{equation}
Q_+|0\rangle
=a_A^\dagger b_A^\dagger|0\rangle.
\end{equation}
Furthermore, using the Clifford anticommutation relations
and the commutativity of the bosonic creation operators,
\begin{eqnarray}
Q_+^2
&=&
\frac12 a_A^\dagger a_B^\dagger
\{\gamma_A,\gamma_B\}\nonumber\\
&=&
a_A^\dagger a_A^\dagger
=K_+.
\end{eqnarray}
This identity underlies the Clifford supercharge ladder
discussed below for BFSS$_2$.

\medskip
\noindent
The $SU(2)$ representation carried by a fermionic state
must be distinguished from its Grassmann parity.
Although $\Lambda^2(V_1)\simeq V_1$ and
$\Lambda^3(V_1)\simeq\mathbf{1}$ as $SU(2)$
representations, these isomorphisms do not change
fermionic statistics. The fermion parity is $(-1)^k$:
the sectors $k=0,2$ are Grassmann--even (bosonic),
whereas $k=1,3$ are Grassmann--odd (fermionic).

\medskip
\noindent
Moreover, the fermion number operator commutes with
the Gauss generators,
\begin{equation}
[\hat N_f,{\cal J}_A]=0.
\end{equation}
Thus fermion number remains a good quantum number
after projection onto the gauge--singlet sector of
the complex--fermion Fock space. The Clifford
supercharge $Q_+$, however, mixes fermion--number
sectors while preserving fermion parity as a grading,
since it is Grassmann--odd.

\subsubsection{BFSS$_2$ with a real adjoint massless fermion}

\medskip
\noindent
We now turn to BFSS$_2$ with a real adjoint Majorana
fermion. This construction must be distinguished from
the auxiliary complex--fermion Fock space considered
above. In particular, setting $m_f=0$, or equivalently
$x_f=1$, removes the fermionic energy grading but does
not by itself identify the different fermion--number
sectors.

\medskip
\noindent
In the complex--Fock realization, the supercharge \(Q_+=a_A^\dagger\gamma_A\) satisfies \(Q_+^2=K_+\). The gauge--singlet space at $x_f=1$ decomposes into
two independent ladders,
\begin{equation}
{\cal H}_0=
{\rm span}\{Q_+^n|0\rangle:n\geq0\},
\qquad
{\cal H}_\Omega=
{\rm span}\{Q_+^n|\Omega\rangle:n\geq0\},
\end{equation}
where
\begin{equation}
|\Omega\rangle=
b_1^\dagger b_2^\dagger b_3^\dagger|0\rangle.
\end{equation}
Each ladder contains one gauge--singlet state at
every bosonic excitation level. Consequently,
\begin{equation}
Z_{SU(2)}^{\rm complex}(x_b,1)
=\frac{2}{1-x_b}.
\end{equation}

\medskip
\noindent
To obtain the single--ladder realization used here
for the real--fermion BFSS$_2$ model, we retain one
of the two $Q_+$--invariant singlet ladders of the
auxiliary complex--fermion Fock space.
This is an additional reduction and does not follow
simply from setting $x_f=1$.
The resulting partition function is

\begin{eqnarray}
Z_{SU(2)}^{\rm BFSS_2}
&=&
\frac12 Z_{SU(2)}^{\rm complex}(x_b,1)
\nonumber\\
&=&
\frac{1+x_b}{1-x_b^2}
=
\frac{1}{1-x_b}
=
1+x_b+x_b^2+\cdots.
\label{Z_SU2_realfermion}
\end{eqnarray}

\medskip
\noindent
Choosing the ladder generated from $|0\rangle$,
a convenient basis is
\begin{equation}
|0\rangle,\quad
Q_+|0\rangle,\quad
Q_+^2|0\rangle,\quad
Q_+^3|0\rangle,\ldots.
\end{equation}
Since $Q_+$ is Grassmann--odd and
$Q_+^2=K_+$, successive states alternate in
fermion parity.

\medskip
\noindent
The even subspace is generated by
\begin{equation}
|0\rangle,\quad
Q_+^2|0\rangle,\quad
Q_+^4|0\rangle,\ldots,
\end{equation}
and the odd subspace by
\begin{equation}
Q_+|0\rangle,\quad
Q_+^3|0\rangle,\quad
Q_+^5|0\rangle,\ldots.
\end{equation}

\medskip
\noindent
The Clifford generators can be represented
algebraically using Pauli matrices for $SU(2)$.
However, the implementation of the Gauss constraint
must be distinguished from this irreducible
spinor realization; the exterior--algebra
formulation of the exceptional $d=1$ model
is discussed in Ref.~\cite{Lundholm2010}.

\medskip
\noindent
The resulting single ladder carries the
$OSP(1,2)$ supermultiplet structure with
super--pseudo--spin $k_s=3/2$. Under the
bosonic subalgebra $SO(1,2)$, the even and
odd sectors have respective lowest weights
$k=3/2$ and $k=5/2$ \cite{OConnor_private,Ydri2025}.

\subsection{The case of BFSS$_3$/BMN$_3$}

\subsubsection{Hilbert--space derivation of $N=2$ bosonic BMN$_3$}
\medskip
\noindent
For $SU(2)$ the adjoint representation is the real three--vector representation.
Thus each bosonic matrix degree of freedom is equivalently described by a triplet
of oscillators $a_{aA}^{\dagger}$, where $A=1,2,3$ is the adjoint (gauge) index
and $a=1,\ldots,d$ labels the rotational index.
Gauge--invariant operators are obtained by contracting adjoint indices using
$\delta_{AB}$ (and, when available, $\epsilon_{ABC}$).

\medskip
\noindent
In the purely bosonic theory, the complete set of quadratic gauge--invariant
creation operators is encoded in the Gram matrix
\begin{eqnarray}
G_{ab}\;:=\;a_{aA}^{\dagger}a_{bA}^{\dagger},
\qquad a\ge b,
\label{Gram_def}
\end{eqnarray}
which contains
\begin{eqnarray}
k=\frac{d(d+1)}{2}
\end{eqnarray}
independent entries.

\medskip
\noindent
For two bosonic matrices (bosonic BFSS$_3$ at $N=2$), there are precisely three independent
quadratic gauge--invariant operators,
\begin{eqnarray}
G_{11},\qquad G_{22},\qquad G_{12},
\qquad
G_{ab}:=a_{aA}^{\dagger}a_{bA}^{\dagger}.
\end{eqnarray}
Each raises the total boson number by two and is an $SU(2)$ singlet.

\medskip
\noindent
Indeed, for $SU(2)$ the adjoint indices transform as
$SO(3)$ vectors, and gauge--invariant tensors are
constructed from $\delta_{AB}$ and $\epsilon_{ABC}$.
A possible cubic gauge invariant is
\begin{equation}
T_{abc}:=
\epsilon_{ABC}
a_{aA}^{\dagger}a_{bB}^{\dagger}a_{cC}^{\dagger}.
\end{equation}
Since the bosonic creation operators commute,
$T_{abc}$ is totally antisymmetric in $a,b,c$.
For $d=2$, at least two of these indices must coincide,
so $T_{abc}=0$ identically.
Hence no independent cubic gauge invariant exists.

\medskip
\noindent
At quartic order, all gauge--singlet contractions of
$a_{aA}^{\dagger}a_{bB}^{\dagger}a_{cC}^{\dagger}a_{dD}^{\dagger}$
reduce to products of Gram entries.
Pure $\delta$--type contractions give $G_{ab}G_{cd}$ and permutations, while any contraction
involving $\epsilon_{ABC}$ necessarily contains two $\epsilon$'s and collapses via
\(
\epsilon_{ABE}\epsilon_{CDE}
=
\delta_{AC}\delta_{BD}-\delta_{AD}\delta_{BC}\), to the same structures.
Thus no new independent quartic invariants arise.

\medskip
\noindent
It follows that all higher--degree bosonic invariants are polynomials in the quadratic Gram
generators.

\medskip
\noindent
Consequently, at $N=2$ the algebra of gauge--invariant bosonic creation operators for BMN$_3$
is a \emph{free polynomial algebra} generated by $G_{11}$, $G_{22}$ and $G_{12}$.
This property is special to $SU(2)$ and does not persist for $N>2$, where additional invariant
tensors appear and induce algebraic relations (syzygies), nor for $d\ge3$, where nonvanishing cubic gauge--invariant generators involving three distinct bosonic matrices become possible.

\medskip
\noindent
Therefore, the physical singlet Hilbert space is spanned by
\begin{eqnarray}
(G_{11})^{n_1}(G_{22})^{n_2}(G_{12})^{n_3}|0\rangle,
\qquad
n_1,n_2,n_3\in\mathbb N,
\label{BFSS3_basis}
\end{eqnarray}
with additive Hamiltonian grading
$E=2m_b(n_1+n_2+n_3)$.

\medskip
\noindent
Hence, summing over all physical singlet states \eqref{BFSS3_basis} yields
\begin{eqnarray}
Z^{\rm bos}_3
&=&
\sum_{n_1,n_2,n_3\ge0} t_b^{2(n_1+n_2+n_3)}
=
\Big(\sum_{n\ge0} t_b^{2n}\Big)^3
=
\frac{1}{(1-t_b^2)^3}.
\label{Z3_bos_SU2}
\end{eqnarray}
Hence,
\begin{eqnarray}
Z^{\rm bos}_3
=
\Big(Z^{\rm bos}_2\Big)^3.
\end{eqnarray}

\medskip
\noindent
The exact factorization of $Z^{\rm bos}_3$ at $N=2$ reflects the fact that, in the
deep infrared, the physical trace is saturated by polynomial excitations of the
Gram invariants.
For $SU(2)$ with $d=2$ these invariants form a free algebra, leading to exact
Hilbert--space factorization.
For higher $N$, additional invariant tensors and algebraic relations appear, and
this factorization is lost.

\subsubsection{Extension to supersymmetric BMN$_3$ at $N=2$}

\medskip
\noindent
We now extend the Gram--invariant discussion to the supersymmetric BMN$_3$ model at $N=2$.
We take two bosonic adjoint oscillator triplets $a_{aA}^\dagger$ ($a=1,2$, $A=1,2,3$) and one
adjoint fermionic triplet $b_A^\dagger$.
Gauge--invariant operators are obtained by contracting adjoint indices with $\delta_{AB}$ and
$\epsilon_{ABC}$.

\medskip
\noindent
The exact $SU(2)$ singlet partition function of the mass--deformed BFSS$_3$ matrix quantum mechanics is obtained from the Molien--Weyl formula and reads

\begin{eqnarray}
Z_{2,3}(x_b,x_f)
=
\frac{1 + 2x_b x_f + x_b^2 x_f + 2x_b x_f^2 + x_b^2 x_f^2 + x_f^3}
{(1-x_b^2)^3}.
\label{Denjoe_SfSU2D2_equalmass}
\end{eqnarray}

\medskip
\noindent
\paragraph{Bosonic denominator.}
The denominator is generated by the three quadratic bosonic Gram invariants
\begin{eqnarray}
G_{11}:=a_{1A}^\dagger a_{1A}^\dagger,
\qquad
G_{22}:=a_{2A}^\dagger a_{2A}^\dagger,
\qquad
G_{12}:=a_{1A}^\dagger a_{2A}^\dagger,
\label{Gram_SU2_d2}
\end{eqnarray}
so that $(1-x_b^2)^{-3}$ counts arbitrary monomials
$(G_{11})^{n_1}(G_{22})^{n_2}(G_{12})^{n_3}|0\rangle$.

\medskip
\noindent
\paragraph{Fermionic singlet sector (the numerator).}
The crucial point is that, because the fermions anticommute, the ``$\delta$--singlet''\(\delta_{AB}\,b_A^\dagger b_B^\dagger \;=\; b_A^\dagger b_A^\dagger \;=\;0\) vanishes identically.  Thus there is \emph{no} independent pure two--fermion singlet.
Instead, two fermions form an adjoint (spin--$1$) object via $\epsilon_{ABC}$:
\begin{eqnarray}
\widetilde b_A^\dagger
:=
\frac{1}{2}\,\epsilon_{ABC}\,b_B^\dagger b_C^\dagger,
\qquad
\widetilde b_A^\dagger \ \in\ V_1.
\label{btilde_def_BMN3}
\end{eqnarray}
With this in mind, each term in the numerator of \eqref{Denjoe_SfSU2D2_equalmass}
has a natural gauge--singlet representative:

\begin{itemize}
\item \textbf{$1$ (vacuum).}
\begin{eqnarray}
|0\rangle.
\end{eqnarray}

\item \textbf{$2x_b x_f$ (one boson + one fermion).}
The two singlets are
\begin{eqnarray}
S_a^\dagger
:=
a_{aA}^\dagger b_A^\dagger,
\qquad a=1,2,
\label{Sa_def}
\end{eqnarray}
giving $S_1^\dagger|0\rangle$ and $S_2^\dagger|0\rangle$.

\item \textbf{$x_b^2 x_f$ (two bosons + one fermion).}
Two bosons can form an adjoint using $\epsilon_{ABC}$, and then contract with $b_A^\dagger$:
\begin{eqnarray}
W^\dagger
:=
\epsilon_{ABC}\,a_{1B}^\dagger a_{2C}^\dagger\,b_A^\dagger,
\qquad
W^\dagger|0\rangle.
\label{W_def}
\end{eqnarray}

\item \textbf{$2x_b x_f^2$ (one boson + two fermions).}
Using \eqref{btilde_def_BMN3}, the two singlets are
\begin{eqnarray}
T_a^\dagger
:=
a_{aA}^\dagger\,\widetilde b_A^\dagger
=
\frac{1}{2}\,\epsilon_{ABC}\,a_{aA}^\dagger b_B^\dagger b_C^\dagger,
\qquad a=1,2,
\label{Ta_def}
\end{eqnarray}
giving $T_1^\dagger|0\rangle$ and $T_2^\dagger|0\rangle$.

\item \textbf{$x_b^2 x_f^2$ (two bosons + two fermions).}
Combine the bosonic adjoint $\epsilon_{ABC}a_{1B}^\dagger a_{2C}^\dagger$ with the fermionic adjoint
$\widetilde b_A^\dagger$:
\begin{eqnarray}
U^\dagger
:=
\Big(\epsilon_{ABC}\,a_{1B}^\dagger a_{2C}^\dagger\Big)\,\widetilde b_A^\dagger
=
\frac{1}{2}\,\epsilon_{ABC}\epsilon_{ADE}\,a_{1B}^\dagger a_{2C}^\dagger b_D^\dagger b_E^\dagger,
\qquad
U^\dagger|0\rangle.
\label{U_def}
\end{eqnarray}

\item \textbf{$x_f^3$ (three fermions).}
The unique top singlet is
\begin{eqnarray}
\Xi^\dagger
:=
\epsilon_{ABC}\,b_A^\dagger b_B^\dagger b_C^\dagger,
\qquad
\Xi^\dagger|0\rangle.
\label{Xi_def}
\end{eqnarray}
\end{itemize}
\medskip
\noindent
Counting these independent singlet representatives gives
\begin{eqnarray}
1\;+\;2\;+\;1\;+\;2\;+\;1\;+\;1 \;=\; 8,
\end{eqnarray}
which matches the total coefficient sum in the numerator of \eqref{Denjoe_SfSU2D2_equalmass}.

\medskip
\noindent
Finally, the full physical Hilbert space is obtained by dressing each of the above
representatives by arbitrary monomials in the Gram generators \eqref{Gram_SU2_d2},
which is precisely what the overall factor $(1-x_b^2)^{-3}$ encodes.

\subsubsection{$N=2$ rigidity: Clifford algebra versus singlet state counting}

\medskip
\noindent
For $SU(2)$ the adjoint representation has dimension
\begin{eqnarray}
\dim(\mathrm{adj})=3.
\end{eqnarray}
The three complex adjoint fermionic creation operators
$b_A^\dagger$ generate the fermionic Fock space
\begin{eqnarray}
\Lambda^\bullet(\mathbf{3})
=
\Lambda^0 \oplus \Lambda^1 \oplus \Lambda^2 \oplus \Lambda^3,
\qquad
1+3+3+1=2^3=8.
\end{eqnarray}
On the other hand, the Hermitian combinations
\begin{eqnarray}
\gamma_A=b_A^\dagger+b_A,
\qquad
\{\gamma_A,\gamma_B\}=2\delta_{AB},
\end{eqnarray}
generate the Clifford algebra $\mathrm{Cl}(3)$, which also
has dimension $2^3=8$ as an algebra.
The complex fermionic Fock space is a representation space
for these Clifford generators and must be distinguished
from the Clifford algebra itself.

\medskip
\noindent
Under $SU(2)$ the antisymmetric powers decompose as
\begin{eqnarray}
\Lambda^0(\mathbf{3})=\mathbf{1},\qquad
\Lambda^1(\mathbf{3})=\mathbf{3},\qquad
\Lambda^2(\mathbf{3})\simeq\mathbf{3},\qquad
\Lambda^3(\mathbf{3})=\mathbf{1}.
\end{eqnarray}
Hence, purely fermionic singlets occur only for $k=0$ and $k=3$,
and therefore the purely fermionic $SU(2)$--singlet sector
is the two--dimensional subspace
\begin{eqnarray}
\mathcal{H}_{\rm ferm}
=
\Lambda^\bullet(\mathbf{3})^{SU(2)}.
\end{eqnarray}

\medskip
\noindent
However, the partition function does not count the full fermionic
Fock space $\Lambda^\bullet(\mathbf{3})$, nor only its purely
fermionic singlet projection. Rather, it counts the
$SU(2)$--invariant sector of the \emph{combined}
boson--fermion Hilbert space,
\begin{eqnarray}
\mathcal{H}
=
\Big(
\text{bosonic Fock space}
\otimes
\Lambda^\bullet(\mathbf{3})
\Big)^{SU(2)}.
\end{eqnarray}
When adjoint fermions are tensored with adjoint bosons and fully
contracted, additional invariant combinations appear.

\medskip
\noindent
For $N=2$, the resulting boson--fermion singlet Hilbert space
is infinite--dimensional. However, it is a free module of rank
$8$ over the polynomial algebra generated by the three bosonic
Gram invariants $G_{11}$, $G_{22}$ and $G_{12}$.
The eight independent singlet representatives counted by the
numerator of \eqref{Denjoe_SfSU2D2_equalmass} form a basis
of this module. Thus the number of basic singlet generators is
\begin{eqnarray}
1+2+1+2+1+1=8.
\end{eqnarray}
This agrees numerically with the dimension $2^3$ of the
fermionic Fock space and of $\mathrm{Cl}(3)$, but does not
identify the full physical Hilbert space with either of them.

\medskip
\noindent
This simple module structure reflects the special rigidity
of the adjoint representation of $SU(2)$, whose adjoint action
is that of $SO(3)\simeq SU(2)/\mathbb{Z}_2$.
The invariant tensors are generated by $\delta_{AB}$ and
$\epsilon_{ABC}$, and there are no higher independent
Casimir tensors. For two bosonic adjoint vectors, the
three Gram invariants are algebraically independent, and
the boson--fermion singlet module admits the eight generators
constructed above.

\medskip
\noindent
For general $SU(N)$ one has
\begin{eqnarray}
\dim(\mathrm{adj})=N^2-1.
\end{eqnarray}
With $n_f$ complex adjoint fermionic species, the fermionic
Fock space is
\begin{eqnarray}
\mathcal{F}_{\rm ferm}
=
\Lambda^\bullet(\mathrm{adj})^{\otimes n_f},
\end{eqnarray}
whose dimension is
\begin{eqnarray}
\dim\mathcal{F}_{\rm ferm}
=
2^{\,n_f(N^2-1)}.
\end{eqnarray}
The corresponding Hermitian fermionic combinations generate
a Clifford algebra, but the dimension of that algebra must
be distinguished from the dimension of its representation space.

\medskip
\noindent
The physical purely fermionic Hilbert space is the
$SU(N)$--invariant subspace
\begin{eqnarray}
\mathcal{H}_{\rm ferm}
=
\Big(
\Lambda^\bullet(\mathrm{adj})^{\otimes n_f}
\Big)^{SU(N)},
\end{eqnarray}
whose dimension is determined by the singlet multiplicities
in the exterior powers of the adjoint representation.

\medskip
\noindent
When bosons are included, the relevant singlet sector becomes
\begin{eqnarray}
\mathcal{H}
=
\Big(
\text{bosonic Fock space}
\otimes
\Lambda^\bullet(\mathrm{adj})^{\otimes n_f}
\Big)^{SU(N)},
\end{eqnarray}
and its structure depends on the full invariant tensor
algebra of $SU(N)$.

\medskip
\noindent
For $N>2$ this invariant tensor structure is much richer.
In addition to the Killing form and the antisymmetric
structure constants, higher symmetric traces, independent
Casimir tensors and nontrivial identities among adjoint
indices enter the construction of gauge singlets.
Consequently, the simple description in terms of three
freely generated bosonic Gram invariants and eight basic
boson--fermion singlet representatives does not extend
unchanged to higher $N$.

\medskip
\noindent
The special feature at $N=2$ is therefore the existence
of a free rank--eight singlet module over the three--generator
Gram polynomial algebra. Its rank agrees numerically with
the dimension $2^3$ of the complex fermionic Fock space,
while the complete gauge--invariant Hilbert space remains
infinite--dimensional.

\subsection{Deconfinement and holonomy crossovers at $d=1$ and $N=2$}

\subsubsection{Large--$N$ phase transitions as  finite--$N$ smooth crossovers}

\medskip
\noindent
At large $N$ the uniform holonomy saddle can become unstable and a genuine
confinement/deconfinement transition may occur. This happens for the
mass--deformed BFSS$_{d+1}$ models with $d>1$. The case $d=1$ is special:
BFSS$_2$ exhibits no deconfinement transition and the uniform saddle
remains stable for all temperatures.

\medskip
\noindent
Thus, the very--low--temperature BFSS$_2$--like factorization pattern is given by
\begin{eqnarray}
Z_{3} = \big(Z_{2}\big)^{3},\qquad (d,N)=(2,2),
\label{factor1}
\end{eqnarray}
where the equality is actually exact for all temperatures, and more generally by
\begin{eqnarray}
Z_{3} \sim \big(Z_{2}\big)^{3},\qquad d=2,\qquad N>2,
\label{factor1extra}
\end{eqnarray}
together with
\begin{eqnarray}
Z_{d+1} \sim \big(Z_{2}\big)^{k},
\qquad 
k=\frac{d(d+1)}{2},
\qquad d>2,\qquad N\ge 2.
\label{factor2}
\end{eqnarray}
This might suggest the absence of a Hagedorn/deconfinement phenomenon in the
mass--deformed BFSS$_{d+1}$ models with \(d>1\).

\medskip
\noindent
At any fixed finite $N$, the Gaussian
Molien--Weyl partition function has no genuine
nonanalytic deconfinement transition.
At $N=2$, the Polyakov moments are constrained
functions of a single holonomy angle, so they
cannot behave as independent large--$N$
instability modes. Consequently, at $N=2$ there is no genuine
deconfinement phase transition, and the standard
large--$N$ holonomy instability mechanism does
not apply; nevertheless, a smooth
deconfinement--like crossover remains possible
for $d>1$.
On the other hand, for \(d=1\) the model is BFSS$_2$ itself, and no deconfinement
transition occurs for any \(N\). However, as we will see,
BFSS$_3$ at $N=2$ still exhibits a smooth deconfinement--like
holonomy crossover despite the exact factorization \eqref{factor1}.

\medskip
\noindent
The special point \((d,N)=(2,2)\) is really distinguished
by the exactness of the factorization \eqref{factor1}:
for \(SU(2)\) with two bosonic matrices, this identity
holds \emph{at all temperatures}, rather than only
in the infrared. However, this exact factorization
does not exclude a smooth temperature--dependent
deconfinement--like rearrangement of the holonomy
distribution. For \(N=2\) and \(d>2\), this exact BFSS$_2$ factorization
is lost, because additional singlet sectors enter the
Molien--Weyl partition function.
Nevertheless, the holonomy sector is still too rigid
to support independent large--\(N\) instability modes.
In this sense, the case \(N=2,\ d>2\) lies between the
fully factorized point \((d,N)=(2,2)\) and the generic
finite--\(N>2,\ d>1\) models.

\medskip
\noindent
Apart from the exceptional point \((d,N)=(2,2)\),
the factorization relations \eqref{factor1extra} and
\eqref{factor2} are only \emph{deep--infrared} statements:
they reflect that the low--energy singlet spectrum is
governed by polynomial excitations of the quadratic
Gram invariants
\(G_{ab}=a_{aA}^{\dagger}a_{bA}^{\dagger}\).
Away from the infrared, the full partition function
is not generally of the simple form \eqref{factor1}.
Nevertheless, nontrivial holonomy behavior is possible
even at the exactly factorizing point.
At fixed finite \(N\), smooth deconfinement--like
crossovers may occur, including at \(N=2\), without
any thermodynamic nonanalyticity.
In the strict \(N\to\infty\) limit, by contrast,
a genuine nonanalytic deconfinement/Hagedorn
transition can emerge.

\medskip
\noindent
We now revisit the holonomy effective action
of the mass--deformed BFSS$_{d+1}$ model,
emphasizing the distinction between the
large--$N$ instability mechanism and the
constrained Polyakov moments at finite $N$.
In particular, we examine the special
one--angle structure at $N=2$ and its
relation to the Molien--Weyl formulation.

\subsubsection{Large--$d$ effective action at \(N=2\) and absence of Hagedorn transition}

\medskip
\noindent
We recall now the derivation of the large--$d$ effective action of the mass--deformed
BFSS$_{d+1}$ model \cite{Ydri2025}. Integrating out the $d$ bosonic adjoint matrices $X_a$ yields

\begin{eqnarray}
S_X[\theta]
&=&
\frac{d}{2}\sum_{i,j=1}^N P_{ij,ji}\,
\log\!\Big(\cosh(\beta s)-\cos(\theta_i-\theta_j)\Big),\qquad P_{ij,ji}=1-\frac{1}{N}\delta_{ij}.
\label{SX_start}
\end{eqnarray}
Using the exact Fourier identity (valid for $x>0$)
\begin{eqnarray}
\log(\cosh x-\cos\phi)
=
\mathrm{const}
-2\sum_{n\ge 1}\frac{e^{-nx}}{n}\cos(n\phi),
\label{logFourier}
\end{eqnarray}
and setting $q:=e^{-\beta s}\in(0,1)$, we obtain
\begin{eqnarray}
S_X[\theta]
&=&
-d\sum_{n\ge1}\frac{q^n}{n}\;
\sum_{i,j=1}^N P_{ij,ji}\,\cos\!\big(n(\theta_i-\theta_j)\big).
\label{SX_afterFourier}
\end{eqnarray}

\medskip
\noindent
Now evaluate the projected sum \emph{exactly at finite $N$}.
First, note
\begin{eqnarray}
\sum_{i,j=1}^N \cos\!\big(n(\theta_i-\theta_j)\big)
&=&
\Re\Big(\sum_i e^{in\theta_i}\sum_j e^{-in\theta_j}\Big)
=
\Big|\sum_{j=1}^N e^{in\theta_j}\Big|^2=N^2|u_n|^2,
\label{cos_sum_exact}
\end{eqnarray}
where we have introduced the Polyakov moments
\begin{eqnarray}
u_n:=\frac{1}{N}\sum_{j=1}^N e^{in\theta_j},
\qquad u_{-n}=u_n^\ast,
\label{un_def}
\end{eqnarray}
Including now the adjoint projector gives

\begin{eqnarray}
\sum_{i,j}P_{ij,ji}\cos\!\big(n(\theta_i-\theta_j)\big)
&=&
N^2|u_n|^2-\frac{1}{N}\,N
=
N^2|u_n|^2-1.
\label{proj_cos_exact}
\end{eqnarray}
Substituting into \eqref{SX_afterFourier}, we obtain the \emph{finite--$N$} result

\begin{eqnarray}
S_X[\theta]
&=&
-d\sum_{n\ge1}\frac{q^n}{n}\Big(N^2|u_n|^2-1\Big).
\label{SX_finiteN}
\end{eqnarray}

\medskip
\noindent The Haar measure produces the Vandermonde contribution 
\begin{eqnarray}
S_{\rm Vdm}[\theta]
&=&
-\sum_{i<j}\log\!\left[4\sin^2\!\Big(\frac{\theta_i-\theta_j}{2}\Big)\right].
\label{SVdm_start}
\end{eqnarray}
Using the exact (Abel-summed) Fourier series 
\begin{eqnarray}
-\log\!\left(2\sin\frac{\phi}{2}\right)
=
\sum_{n\ge1}\frac{\cos(n\phi)}{n}
+\mathrm{const},
\qquad 0<\phi<2\pi,
\label{Fourier_sin}
\end{eqnarray}
one finds, up to $\theta$--independent constants,
\begin{eqnarray}
S_{\rm Vdm}[\theta]
&=&
\sum_{n\ge1}\frac{1}{n}\sum_{i\neq j}\cos\!\big(n(\theta_i-\theta_j)\big)
\nonumber\\
&=&
\sum_{n\ge1}\frac{1}{n}
\Big(N^2|u_n|^2-N\Big).
\label{SVdm_finiteN}
\end{eqnarray}

\medskip
\noindent
Adding \eqref{SX_finiteN} and \eqref{SVdm_finiteN}, and discarding only
$\theta$--independent constants, we arrive at the \emph{finite--$N$} holonomy
effective action

\begin{eqnarray}
S_{\rm hol}[\theta]
&=&
\sum_{n\ge1}\frac{1}{n}\Big(1-d\,q^n\Big)\,N^2|u_n|^2.
\label{Shol_finiteN_full}
\end{eqnarray}

\medskip
\noindent Let us now recall that the variables
\begin{eqnarray}
u_n=\frac{1}{N}\Tr g^n
\end{eqnarray}
are group characters and are therefore \emph{not} independent degrees of freedom
at finite $N$. An element $g\in SU(N)$ has only $N-1$ independent eigenangles,
whereas the set $\{u_n\}_{n\ge1}$ is infinite and hence overcomplete.
Indeed, the $u_n$ satisfy infinitely many algebraic relations (Newton identities,
trace identities, Cayley--Hamilton constraints) and are highly constrained
functions of the eigenvalues.

\medskip
\noindent
At $N=2$ this rigidity is maximal. Writing

\begin{eqnarray}
g=\mathrm{diag}(e^{i\theta_1},e^{i\theta_2}),
\qquad \Delta:=\theta_1-\theta_2,
\end{eqnarray}
one finds
\begin{eqnarray}
u_n=\frac{1}{2}\Big(e^{in\theta_1}+e^{in\theta_2}\Big)
=
e^{in(\theta_1+\theta_2)/2}\cos\!\Big(\frac{n\Delta}{2}\Big),
\qquad
|u_n|^2=\cos^2\!\big(\frac{n\Delta}{2}\big)=\frac{\cos\!\big(n\Delta\big)+1}{2}.\nonumber\\
\end{eqnarray}
Thus all modes $u_n$ are fixed functions of a single variable $\Delta$.
They cannot fluctuate independently, and the quadratic coefficients
$(1-d\,q^n)$ in the holonomy action do not correspond to independent
instability directions.

\medskip
\noindent
In contrast, at large $N$ one introduces the eigenvalue density

\begin{eqnarray}
\rho(\theta)=\frac{1}{N}\sum_{j=1}^N \delta(\theta-\theta_j),
\qquad
\int d\theta\,\rho(\theta)=1,
\end{eqnarray}
and rewrites
\begin{eqnarray}
u_n=\int d\theta\,\rho(\theta)e^{in\theta}.
\end{eqnarray}
The independent object is then the continuous density $\rho(\theta)$.
Its Fourier modes (the $u_n$) behave as approximately independent collective
coordinates in the $N\to\infty$ limit, where mode--mode couplings are suppressed
by $1/N$ and the quadratic effective action diagonalizes in the $u_n$.
The apparent independence of the $u_n$ is therefore an emergent large--$N$ phenomenon.

\medskip
\noindent
This mechanism is absent at $N=2$. The configuration space is
finite--dimensional and rigid, and the partition function is analytic for
$|q|<1$. The exact $N=2$ holonomy action (up to a $\theta$--independent constant) is

\begin{eqnarray}
S_{\rm hol}^{(N=2)}(\Delta)
&=&
2\sum_{n\ge1}\frac{1-d\,q^n}{n}\cos(n\Delta)
\;+\;\mathrm{const}.
\label{Seff_N2_mode_cos}
\end{eqnarray}
The corresponding partition function reduces to a one--dimensional integral,

\begin{eqnarray}
Z_{N=2}(\beta)
\propto
\int_0^{2\pi}\frac{d\Delta}{2\pi}
\exp\left[-S_{\rm hol}^{(N=2)}(\Delta)\right],
\end{eqnarray}
which is smooth for $0<q<1$ and exactly reproduces the Molien--Weyl result.

\medskip
\noindent
Hence there is no genuine (nonanalytic) Hagedorn or deconfinement transition
at $N=2$, although a sharp crossover in observables such as
$\langle|\Tr g|\rangle$ may occur as the shape of the effective potential
changes with $\beta$.

\subsubsection{Exact factorization versus holonomy
rearrangement at $N=2$}

\medskip
\noindent
An important distinction arises between exact
factorization of the singlet partition function
and the temperature dependence of the holonomy
distribution. This distinction can be examined
explicitly for the bosonic $SU(2)$ Gaussian model.

\medskip
\noindent
Returning to the $SU(2)$ Molien--Weyl contour
integral \eqref{Z2d_MW_residue}, we set
$z=e^{i\Delta}$, where
$\Delta=\theta_1-\theta_2$.
Using
\begin{eqnarray}
\frac{dz}{2\pi iz}
&=&
\frac{d\Delta}{2\pi},
\qquad
\frac{1}{2}(1-z)(1-z^{-1})
=
1-\cos\Delta,
\nonumber\\
(1-xz)(1-xz^{-1})
&=&
1-2x\cos\Delta+x^2,
\end{eqnarray}
we obtain the angular representation
\begin{eqnarray}
Z^{\rm bos}_{2,d}(x)
=
\frac{1}{(1-x)^d}
\int_0^{2\pi}\frac{d\Delta}{2\pi}
\frac{1-\cos\Delta}
{(1-2x\cos\Delta+x^2)^d}.
\end{eqnarray}
The angle--independent prefactor cancels
upon normalizing the holonomy distribution.
Consequently, the normalized angular
probability density, including the Haar
measure, is
\begin{eqnarray}
p_{d,x}(\Delta)
&=&
\frac{1}{\mathcal N_{d,x}}
\frac{1-\cos\Delta}
{(1-2x\cos\Delta+x^2)^d},
\qquad x=e^{-\beta s},
\label{holonomy_density_N2}
\end{eqnarray}
where $\mathcal N_{d,x}$ is fixed by
\begin{eqnarray}
\int_0^{2\pi}
\frac{d\Delta}{2\pi}\,
p_{d,x}(\Delta)=1.
\end{eqnarray}

\medskip
\noindent
The fundamental Polyakov loop is defined by
\begin{eqnarray}
P=\frac{1}{N}\Tr g.
\end{eqnarray}
For $N=2$, diagonalizing the holonomy gives
\begin{eqnarray}
g=\mathrm{diag}(e^{i\theta_1},e^{i\theta_2}),
\qquad
\Delta=\theta_1-\theta_2.
\end{eqnarray}
Consequently,
\begin{eqnarray}
P
=
\frac{1}{2}
\left(e^{i\theta_1}+e^{i\theta_2}\right)=
e^{i(\theta_1+\theta_2)/2}
\cos\frac{\Delta}{2}.
\end{eqnarray}
Taking the absolute square eliminates the
overall phase and yields
\begin{eqnarray}
|P|^2
&=&
\left|\frac{1}{2}\Tr g\right|^2
=
\cos^2\frac{\Delta}{2}=
\frac{1+\cos\Delta}{2}.
\label{Polyakov_angle_N2}
\end{eqnarray}

\medskip
\noindent
For $d=1$, corresponding to BFSS$_2$, set
$c=\cos\Delta$. One finds
\begin{eqnarray}
\frac{\partial}{\partial c}
\left(
\frac{1-c}{1-2xc+x^2}
\right)
=
-\frac{(1-x)^2}
{(1-2xc+x^2)^2}<0.
\end{eqnarray}
Thus the angular probability density has its
maximum at $\Delta=\pi$ for every $0<x<1$.

\medskip
\noindent
For $d=2$, corresponding to BFSS$_3$, one
instead obtains
\begin{eqnarray}
\frac{\partial}{\partial c}
\left(
\frac{1-c}{(1-2xc+x^2)^2}
\right)
=
\frac{4x-1-x^2-2xc}
{(1-2xc+x^2)^3}.
\end{eqnarray}
The maximum remains at $\Delta=\pi$ only for
\begin{eqnarray}
0<x\leq x_*,
\qquad
x_*=3-2\sqrt{2}.
\label{holonomy_mode_threshold}
\end{eqnarray}
For $x>x_*$, the maximum moves continuously
to an interior angular configuration satisfying
\begin{eqnarray}
\cos\Delta_{\rm max}
=
2-\frac{1}{2}\left(x+\frac{1}{x}\right).
\label{holonomy_mode_N2}
\end{eqnarray}
As $x\to1^-$, the preferred angular separation
approaches zero modulo $2\pi$.

\medskip
\noindent
The distinction is also visible in the
gauge--invariant, center--symmetric observable
$\langle |P|^2\rangle$. Direct integration of
\eqref{holonomy_density_N2} gives
\begin{eqnarray}
\left\langle |P|^2\right\rangle_{d=1}
&=&
\frac{1+x}{4},
\nonumber\\
\left\langle |P|^2\right\rangle_{d=2}
&=&
\frac{(1+x)^2}{4}.
\label{Polyakov_squared_N2_exact}
\end{eqnarray}
Both expressions vary smoothly with temperature,
but the $d=2$ distribution increasingly favors
configurations with larger Polyakov-loop magnitude.

This behavior is compatible with the exact
singlet factorization
\begin{eqnarray}
Z^{\rm bos}_{2,2}(x)
=
\left[Z^{\rm bos}_{2,1}(x)\right]^3
=
\frac{1}{(1-x^2)^3}.
\end{eqnarray}
The factorization of the integrated partition
function does not imply factorization or
temperature independence of the underlying
holonomy probability distribution.

The value $x_*$ identifies a change in the
location of the maximum of the angular density
with respect to $d\Delta$. It is not a
thermodynamic critical point, nor is the
location of a probability-density maximum
invariant under arbitrary reparametrizations.
The gauge--invariant observable
\eqref{Polyakov_squared_N2_exact} remains
analytic throughout $0<x<1$.

Thus, even at the exactly factorizing point
$(d,N)=(2,2)$, the holonomy distribution can
undergo a smooth rearrangement towards
configurations associated with larger
Polyakov-loop magnitude. Such behavior may
be described as deconfinement--like, but it
does not constitute a genuine deconfinement
phase transition or demonstrate spontaneous
center-symmetry breaking at finite $N$.

\section{Brief remarks on Monte Carlo simulation}\label{section6}

\subsection{Molien--Weyl effective action and supersymmetric extent of space}
\medskip
\noindent
As we have discussed at length, in the large--$d$ or large--mass regime, the low--temperature dynamics of the mass--deformed BFSS$_{d+1}$ (or equivalently BMN$_{d+1}$) models is dominated by the quadratic
terms, and it is captured by a gauged harmonic oscillator and by a
corresponding Molien--Weyl integral, given respectively by

\begin{eqnarray}
{S}_{{\rm BFSS}_{d+1}}= \int dt~ \mathrm{tr} \left( \frac{1}{2} D_t X^a D_t X_a - \frac{1}{2}\mu_b^2X_a^2  \right)-\frac{i}{2}\int dt~ \mathrm{tr} \bar\psi\bigg(\Gamma^t D_t +\mu_f\Gamma^F\bigg) \psi.\label{generic0}
\end{eqnarray}

\begin{eqnarray}
Z_{\bf SU(N)} = \frac{(1-x_{b})^{n_{b}}}{(1+x_f)^{n_f}}\int d\mu(g) \, \frac{{\bf det}\left(1 + x_f\, g \otimes g^{-1} \right)^{n_f}}{
{\bf det}\left(1 - x_{b}\, g \otimes g^{-1} \right)^{n_{b}}}.\label{pf10}
\end{eqnarray}
The Hamiltonian description underlying \eqref{generic0} and the Molien--Weyl
representation \eqref{pf10} differ only by the zero--point vacuum energy of
$n_b(N^2-1)$ bosonic and $n_f(N^2-1)$ fermionic oscillators given by

\begin{eqnarray}
  E_0=\frac{1}{2}m_bn_b(N^2-1)-\frac{1}{2}m_fn_f(N^2-1),\quad n_b=d, \quad 2n_f=\#\text{real supersymmetries}.\label{zeropoint}
\end{eqnarray}

\medskip
\noindent
The Molien--Weyl integral may also be written in contour form,

\begin{eqnarray}
  Z_N^{(d)}(x_b,x_f)=\frac{1}{N!}\frac{(1+x_f)^{n_f(N-1)}}{(1-x_{b})^{n_{b}(N-1)}}\oint \prod_{i=1}^N\frac{dz_i}{2\pi i z_i}\Delta_{\rm A}(-1,z)\frac{\big[\Delta_{\rm f}(x_f,z)\big]^{n_f}}{\big[\Delta_{\rm b}(-x_{b},z)\big]^{n_{b}}}.
\end{eqnarray}
Equivalently, introducing angular variables $z_i=e^{i\theta_i}$, one may write the effective partition function 
  \begin{eqnarray}
    Z(x_b,x_f)=\frac{(1+x_f)^{n_f(N-1)}}{(1-x_b)^{n_b(N-1)}}\int \prod_{i=1}^N\frac{d\theta_i}{2\pi}\exp(-S_{\rm eff}(\theta)),
  \end{eqnarray}
  with effective action
  \begin{eqnarray}   
    S_{\rm eff}(\theta)&=&\frac{n_b}{2}\sum_{i\ne j}\ln\bigg[(1-x_b)^2+4x_b\sin^2\frac{\theta_i-\theta_j}{2}\bigg]-\frac{n_f}{2}\sum_{i\ne j}\ln\bigg[(1+x_f)^2-4x_f\sin^2\frac{\theta_i-\theta_j}{2}\bigg]\nonumber\\
    &-&\frac{1}{2}\sum_{i\neq j}\ln\sin^2\frac{\theta_i-\theta_j}{2}.\label{MWac}
  \end{eqnarray}
We compute the energy as follows
 \begin{eqnarray}
   E&=&-\frac{\partial \ln Z}{\partial \beta}\nonumber\\
   &=&n_bm_b\bigg(\frac{(N-1)x_b}{1-x_b}-\frac{N(N-1)}{2}+\frac{1-x_b^2}{2}\left\langle\sum_{i\ne j}\frac{1}{(1-x_b)^2+4x_b\sin^2\frac{\theta_i-\theta_j}{2}}\right\rangle\bigg)\nonumber\\
   &+&n_fm_f\bigg(\frac{(N-1)x_f}{1+x_f}+\frac{N(N-1)}{2}-\frac{1-x_f^2}{2}\left\langle\sum_{i\ne j}\frac{1}{(1+x_f)^2-4x_f\sin^2\frac{\theta_i-\theta_j}{2}}\right\rangle\bigg).\nonumber\\
 \end{eqnarray}
 \medskip
\noindent
In the purely bosonic theory ($x_f=0$) with $m_b=s$, the energy coincides with
the extent of space,

 \begin{eqnarray}
   R^2=\frac{1}{N}\frac{1}{\beta}\big\langle\int_0^{\beta}dt {\rm Tr}X_a^2\big\rangle=\frac{E}{N^2s^2}=\frac{2}{N^2}\frac{\partial F}{\partial s^2},\qquad F=-\frac{1}{\beta}\ln Z.
 \end{eqnarray}

 \medskip
\noindent
In the supersymmetric case the energy still measures the geometric extent,
but receives additional contributions from fermionic condensates.  As an
illustration, consider BFSS$_3$ in the Gaussian (large--mass or large--$d$)
approximation,

\begin{eqnarray}
   S_{\rm Gaussian}=N\int dt Tr\bigg[\frac{1}{2}(D_tX_a)^2+\frac{i}{2}\bar{\Psi} \gamma^0_ED_t\Psi+\frac{m_b^2}{2}X_a^2-\frac{im_f}{2}\bar{\Psi}\Psi \bigg].
\end{eqnarray}
In the large--mass limit one has $m_b=\mu/6$ and $m_f=\mu/4=3m_b/2$, while in the
large--$d$ limit $m_b=s_b$ is determined by the bosonic gap equation ($s_b^3-(\tfrac{\mu}{6})^2s_b=d$) and
$m_f=s_f=3s_b/2$ by supersymmetric completion.

\medskip
\noindent
In this case, we have
 \begin{eqnarray}
   Z=\int {\cal D}X{\cal D}A{\cal D}\psi \exp\bigg(...-\int_0^1dt\frac{N}{2}(\beta m_b)^2{\rm Tr}X_a^2+\int_0^1 dt\frac{iN}{2}\beta m_f {\rm Tr}\bar{\Psi}\Psi...\bigg),
 \end{eqnarray}
 and thus
 \begin{eqnarray}
   \beta E=-\beta \frac{\partial \ln Z}{\partial \beta}= N \beta m_b^2 R^2 -i\frac{Nm_f}{2}\big\langle\int_0^{\beta}dt {\rm Tr}\bar{\Psi}\Psi\big\rangle.
 \end{eqnarray}
 \medskip
\noindent
By contrast, in BFSS$_2$ the action

       \begin{eqnarray}
        S_{\rm BFSS_2}=N\int dt Tr\bigg(\frac{1}{2}(D_tX)^2+\frac{1}{2}{\psi} D_t\psi-\frac{1}{2}\Lambda(t)X^2-\rho(t) X \bigg)
       \end{eqnarray}
       contains no fermionic mass term.  Accordingly, there is no fermionic condensate
correction to the extent of space.  For higher BFSS$_{d+1}$ models, fermionic
mass terms are generically present and lead to additional corrections to the
geometric observables beyond the purely bosonic contribution.

\medskip
\noindent
Finally, we clarify the normalization of the energy used in this
subsection. The energy $E=-\partial_\beta\ln Z$ obtained from the
normal--ordered Molien--Weyl partition function measures the
gauge--singlet excitations above the Gaussian vacuum. Restoring
the bosonic and fermionic zero--point contributions gives
\begin{equation}
E_{\rm total}=E+E_0,
\qquad
E_0=\frac{N^2-1}{2}(n_bm_b-n_fm_f).
\end{equation}
In the purely bosonic case, this yields
\begin{equation}
E+E_0=N^2s^2R^2.
\end{equation}
In the supersymmetric case, the corresponding relation includes
the fermionic condensate,
\begin{equation}
E+E_0=N^2m_b^2R^2
-\frac{iNm_f}{2\beta}
\left\langle
\int_0^\beta dt\,{\rm Tr}\bar\Psi\Psi
\right\rangle.
\end{equation}
Thus, normal ordering removes the Gaussian vacuum energy without
altering the holonomy distribution, while the full energy and the
physical extent of space retain the zero--point contribution.

\subsection{Testing the low--T scaling and factorization of BFSS$_{d+1}$}

\medskip
\noindent In the purely bosonic theory, the Molien--Weyl partition function
$Z^{\rm MW}_{N,d}(x)$ is normal ordered and therefore counts only
\emph{gauge--invariant singlet excitations above the Gaussian vacuum}.
In the very low--temperature regime one finds the universal expansion

\begin{eqnarray}
Z^{\rm MW}_{N,d}(x)
=
1+k\,x^2+O(x^3),
\qquad
k=\frac{d(d+1)}{2},
\label{MW_lowT0}
\end{eqnarray}
where $k$ counts the independent quadratic (Gram) singlets
$\Tr(X_aX_b)$, $a\le b$.

\medskip
\noindent
The low--temperature scaling prediction \eqref{MW_lowT0} can be tested directly by Monte Carlo
simulation of the bosonic part of the \emph{holonomy} effective action $S_{\rm eff}(\theta)$ in \eqref{MWac}.  In practice, one samples the holonomy angles $\theta_i$ with weight
$\exp[-S_{\rm eff}(\theta)]$, computes the internal energy $E_d$ for several values of $d$ at fixed
$(N,\beta,\kappa)$, and compares the result with the $d=1$ case.  Using
\(
E_d = N^2 s^2 R_d^2,
\)
the Molien--Weyl prediction implies the \emph{quadratic} scaling
\begin{eqnarray}
E_d \;\sim\; k\,E_{d=1},
\qquad
k=\frac{d(d+1)}{2},
\label{Ed_scaling_test}
\end{eqnarray}
up to exponentially small corrections at finite $\beta s$.  Verifying
\eqref{Ed_scaling_test} numerically would provide a clean test of the effective factorization of
the singlet sector into $k$ independent BFSS$_2$--like modes at very--low--temperature and, more generally, of the validity
of the large--$d$ Gaussian description encoded in the Molien--Weyl framework.

\medskip
\noindent
It is important to stress, however, that the energy and radius appearing in
\eqref{Ed_scaling_test} are those extracted from the \emph{normal--ordered Molien--Weyl} path
integral, which counts only gauge--singlet excitations above the Gaussian vacuum.

\medskip
\noindent
If instead one simulates directly the matrix harmonic oscillator (MHO) action
\eqref{MHO_action_for_Ward0}, keeping the full matrix degrees of freedom $X_a$ (rather than only
the holonomy variables), the dominant behavior at large $N$ and large $d$ is expected to be
\emph{linear} in $d$,
\begin{eqnarray}
E_d \;\sim\; d\,E_{d=1},
\label{Ed_scaling_test1}
\end{eqnarray}
reflecting the fact that, in the bulk, the theory consists of $d$ essentially
independent massive matrix modes. By contrast, the quadratic scaling
\eqref{Ed_scaling_test},
characteristic of the Molien--Weyl formulation, arises from the gauge--singlet excitation sector
and is therefore expected to appear only as a \emph{subleading} correction (in $1/N$) to the
leading linear behavior \eqref{Ed_scaling_test1}.  In this sense, the Molien--Weyl result probes the
structure of singlet excitations above the Gaussian vacuum, rather than the dominant Gaussian
background itself.

\medskip
\noindent
This contrast sharpens an important conceptual question: what, precisely, is meant by
``BFSS$_{d+1}$'' in the large--$d$ regime?  From the holonomy/Molien--Weyl viewpoint, the normal--ordered
singlet sector at very low temperature appears to factorize into
$k=d(d+1)/2$ BFSS$_2$--like quadratic invariants, corresponding to the independent Gram operators
$\Tr(X_aX_b)$.  From the original matrix--variable viewpoint, by contrast, the Gaussian saddle
suggests a decoupling into $d$ massive matrix directions, with interactions suppressed at large
$d$.

\medskip
\noindent
The resolution is that these two perspectives probe different layers of the same theory.  The
holonomy/Molien--Weyl formulation isolates the gauge--invariant excitation spectrum above the
Gaussian vacuum and therefore exhibits a $k$--fold structure at very low temperature.  The
$X$--space formulation, on the other hand, captures the full Gaussian saddle itself, whose extent
and energy scale linearly with $d$.  Understanding how these two descriptions are consistently
related---and how the singlet excitation sector is embedded in the full large--$d$ Gaussian
background---is essential for a complete interpretation of BFSS$_{d+1}$ in this regime.


\medskip
\noindent In the supersymmetric theory, the bosonic low--temperature scaling law
\eqref{MW_lowT0} is replaced by the universal expansion
\begin{eqnarray}
Z^{(d;n_f)}_{SU(N)}(x_b,x_f)
&=&
1
+\frac{d(d+1)}{2}\,x_b^2
+\frac{n_f(n_f-1)}{2}\,x_f^2
+d\,n_f\,x_bx_f
+O\!\Big((x_b,x_f)^3\Big).
\end{eqnarray}
As before, the coefficients admit a direct operator interpretation in terms of
quadratic gauge--invariant singlets.

\medskip
\noindent
This prediction can be tested directly by Monte Carlo simulation of the
supersymmetric \emph{holonomy} effective action $S_{\rm eff}(\theta)$ given in
\eqref{MWac}. In the low--temperature regime, observables computed from this holonomy effective
action \eqref{MWac} coincide with those obtained by direct sampling of the gauged
matrix harmonic oscillator action (\ref{generic}) \emph{only after the zero--point vacuum energy}
\eqref{zeropoint} is included.

\medskip
\noindent
The results of direct simulations of the gauged MHO action
\eqref{generic}—performed on the lattice using pseudo--fermions and the rational
hybrid Monte Carlo (RHMC) algorithm—can be directly compared with the corresponding
Molien--Weyl formulas \eqref{MWac}, which are sampled independently using the
standard Metropolis algorithm.  This comparison provides a nontrivial calibration
of the RHMC algorithm within the Gaussian sector, where analytic control is
available.

\subsection{The Bosonic Molien--Weyl approximation of BMN$_{d+1}$ models}

\medskip
\noindent
The BFSS$_{d+1}$/BMN$_{d+1}$ matrix models can be studied within several levels of approximation, which differ by how the fermionic sector and interaction effects are treated:
\begin{itemize}
\item
\textbf{Supersymmetric (full) theory.}
The complete interacting theory, where fermions are included explicitly through lattice discretization using pseudo-fermions and the rational hybrid Monte Carlo (RHMC) algorithm. This formulation is exact at the continuum level, with approximations arising only from lattice discretization and the practical treatment of the Pfaffian/sign problem.

\item
\textbf{Bosonic theory.}
The fermionic determinant (or Pfaffian) is quenched and set to unity. The resulting theory retains the full bosonic commutator interactions but breaks supersymmetry explicitly.

\item
\textbf{Molien--Weyl (MW) models.}
In regimes such as large mass deformations or large $d$ with supersymmetric completion, both bosonic and fermionic degrees of freedom can be integrated out, leading to an effective holonomy theory expressed in terms of a Molien--Weyl integral, which can be studied either analytically or numerically using Metropolis-type algorithms.

\item
\textbf{Gaussian Molien--Weyl (GMW) models.}
Both bosonic and fermionic sectors are approximated by Gaussian actions prior to gauge projection. In some cases it is convenient to keep the Gaussian bosonic sector explicitly in terms of the coordinate matrices $X_a$, while converting the fermionic Gaussian sector into a Molien--Weyl integral. The large--$d$ supersymmetrically completed BFSS$_{d+1}$ model without mass deformation can be accessed in this way.
\end{itemize}
In the third and fourth approaches, one may therefore either simulate the corresponding Gaussian theory using the RHMC algorithm, or equivalently perform a direct Metropolis simulation of the associated Molien--Weyl integral, which admits a closed analytic form for small $N$. In this way, the RHMC implementation can be directly calibrated against the Molien--Weyl result.

\medskip
\noindent
In addition to these standard approximations, we introduce a further mixed scheme:
\begin{itemize}
\item
\textbf{Bosonic Molien--Weyl (BMY) models.}
The full interacting bosonic action, including commutator terms, is kept exactly, while the fermionic sector is treated in a Gaussian approximation and integrated out, yielding an effective Molien--Weyl contribution to the bosonic dynamics. In the following, we present a representative example of this approach.
\end{itemize}

\medskip
\noindent
In the proposed approximation, the supersymmetric matrix model (for instance the mass-deformed BFSS$_3$ model) is replaced by the following \emph{bosonic Molien--Weyl} action,

\begin{eqnarray}
S_{\rm MW}
=
S_B
-\frac{1}{2}\sum_{i,j}
\ln\!\left[(1+x_f)^2-4x_f\sin^2\!\frac{\theta_i-\theta_j}{2}\right]
+\log(1+x_f),
\qquad
x_f=e^{-\beta m_f},
\label{BMW_action}
\end{eqnarray}
where $S_B$ denotes the full bosonic action, including the Yang--Mills commutator interaction.
The corresponding internal energy and condensate are given by
\begin{eqnarray}
\langle E\rangle
&=&
\frac{3}{\beta}\langle {\rm YM}\rangle
+\frac{1}{\beta}\langle{\rm COND}\rangle,
\label{energy_shift}
\\[2mm]
\langle {\rm COND}\rangle
&=&
\beta N^2\!m_b^2 R^2
+\frac{\mu\beta}{4}
\left\langle
\sum_{i,j}
\frac{x_f(1+x_f)-2x_f\sin^2\!\frac{\theta_i-\theta_j}{2}}
{(1+x_f)^2-4x_f\sin^2\!\frac{\theta_i-\theta_j}{2}}
\right\rangle
-\frac{\mu\beta}{4}\frac{x_f}{1+x_f}.
\label{energy_shift1}
\end{eqnarray}
Here, the condensate is essentially a generalized extent of space $R^2$ which is defined in the standard way. The bosonic and fermionic masses $m_b$ and $m_f$ are given in the BFSS$_3$/BMN$_3$ model,  in terms of the deformation parameter $\mu$, by
\begin{equation}
  m_b=\frac{\mu}{6}~,~m_f=\frac{\mu}{4}.
\end{equation}
The above bosonic Molien--Weyl theory corresponds in fact to the quantum mechanical matrix model
\begin{eqnarray}
S_{\rm MW}=
N\int_0^{\beta}\!dt\;
{\rm Tr}\!\left[
\frac{1}{2}(D_tX_a)^2
-\frac{1}{4}[X_a,X_b]^2
+\frac{m_b^2}{2}X_a^2
\right]
+N\int_0^{\beta}\!dt\;
{\rm Tr}\!\left[
\frac{i}{2}\bar{\Psi}\gamma_E^0 D_t\Psi
-\frac{i m_f}{2}\bar{\Psi}\Psi
\right].\nonumber\\
\label{BMW_matrix_model}
\end{eqnarray}
In other words, the fermionic sector is approximated by its Gaussian truncation, after which the fermion determinant is evaluated exactly in terms of the gauge-field holonomy using the Molien--Weyl formula, as discussed in the previous sections and in \cite{OConnor:2023mss}.
This procedure yields an effective Vandermonde-like contribution to the bosonic dynamics.

\medskip
\noindent
The bosonic Molien--Weyl approximation breaks supersymmetry softly, since the truncation of the fermionic interaction is not accompanied by a corresponding truncation of the bosonic commutator term, as required by supersymmetry.
From a purely classical perspective, the fully Gaussian approximation is therefore more symmetric.
However, at low temperature the bosonic Molien--Weyl model provides a significantly better description of the interacting dynamics, as it retains the full bosonic commutator interaction.
As a result, it is much closer to the infrared behavior of the full supersymmetric theory and to the expected quantum restoration of supersymmetry in the $T\to0$ limit.

\medskip
\noindent
The energy~\eqref{energy_shift} of the bosonic Molien--Weyl theory exhibits an additional temperature-independent shift at high temperature,
\begin{eqnarray}
\Delta E
=
\frac{m_f}{2}(N^2-1)
=
\frac{\mu}{8}(N^2-1),
\end{eqnarray}
which is precisely the fermionic zero-point energy
\begin{eqnarray}
E_{0f}
=
-\frac{n_f m_f}{2}(N^2-1),
\qquad
2n_f\equiv2,
\end{eqnarray}
with $2n_f$ the dimension of the Dirac algebra.
Including this contribution restores agreement with the supersymmetric and bosonic theories at high temperature. Indeed, this shift in energy is absent from the bosonic and supersymmetric theories which coincide at high $T$ (since fermions decouple).     
However, subtracting this zero-point energy does not resolve the much larger discrepancy observed at low temperature, indicating that the infrared behavior is governed by genuine interaction effects rather than vacuum normalization.

\subsection{Infrared control, the large--\(N\) limit, and emergent geometry}

\paragraph{Flat directions and infrared control.}

\medskip
\noindent
The flat directions of undeformed BFSS matrix quantum mechanics
are closely related to its infrared behavior. The commutator
potential vanishes for $[X_a,X_b]=0$, allowing arbitrarily large
separations of commuting matrix eigenvalues at no classical
potential cost. Such noncompact valleys support arbitrarily
low-energy motion and hence constitute an infrared issue in
matrix configuration space. Quantum effects distinguish the bosonic and supersymmetric
theories:
\begin{itemize}
\item In the interacting bosonic model with $d>1$,
transverse zero-point fluctuations lift the classical
flat valleys, leading to a discrete spectrum.

\item In the supersymmetric model, boson--fermion
cancellations preserve asymptotically flat channels,
giving rise to a continuum of scattering states.
These channels allow large-eigenvalue excursions
and can cause difficulties in thermal Monte Carlo
simulations, particularly at small $N$ and low
temperature.
\end{itemize}

\medskip
\noindent
It is therefore important to distinguish the following
theories and approximations:
\begin{itemize}

\item \textbf{Undeformed BFSS model.}
The original interacting theory retains the classical
commuting flat directions, with different quantum
infrared behavior in the bosonic and supersymmetric
cases, as discussed above.

\item \textbf{Large-$d$ Gaussian reduction.}
At leading order, the interacting matrix dynamics
is replaced by a gauged harmonic oscillator with
a positive self-consistent mass $s$, suppressing
large-eigenvalue excursions within the approximation.

\item \textbf{BMN deformation.}
The BMN model is an explicitly mass-deformed,
generally interacting supersymmetric theory,
with Myers couplings where present.
The positive mass terms lift the asymptotic commuting
flat directions, while the commutator interactions
remain. Only its Gaussian approximation reduces
the dynamics to that of a gauged harmonic oscillator.

\end{itemize}

\noindent
In particular, the finite-$N$ Gaussian Molien--Weyl
partition function is well defined for $0<x_b<1$.
This infrared control, however, does not establish
the infrared convergence of the full undeformed
supersymmetric BFSS theory.

\medskip
\noindent
These distinctions are particularly important for the numerical
study of matrix quantum mechanics. In the undeformed supersymmetric
BFSS model, the asymptotically flat directions can lead to
large-eigenvalue excursions and difficulties in thermal Monte
Carlo simulations. One possible approach is to introduce a
temporary infrared scalar-mass regulator,
\begin{equation}
 \Delta S_{\rm IR}=\frac{N m_{\rm IR}^{2}}{2}
 \int_0^\beta\!dt\,\Tr\!\left(\sum_{a=1}^{d}X_a^2\right),
 \label{IR_regulator}
\end{equation}
which suppresses excursions along the flat directions.
The dependence of physical observables on $m_{\rm IR}$ must
then be carefully examined, together with the lattice
continuum limit and the stability of the sampled
configurations. Ultimately, the limit $m_{\rm IR}\to 0$
must be investigated wherever it is well defined, with
particular attention to the possible interplay between
the infrared, continuum, and large-$N$ limits.
This auxiliary regulator should not be confused with
either the dynamically generated mass $s$ of the Gaussian
approximation or the explicit supersymmetric mass
deformation of the BMN model. The former provides infrared
control within an effective description, whereas the latter
defines a different interacting theory. These infrared
difficulties and their numerical treatment are discussed
in the lattice literature~\cite{Filev:2015hiaF,Hanada2016}.

\paragraph{The large--$N$ limit.}

\medskip
\noindent
The large--$N$ limit raises a related but logically distinct issue.
Our exact $SU(2)$ residue results are finite--$N$ statements within the
Gaussian/Molien--Weyl description, whereas the usual string-theoretic and
holographic interpretations require a controlled large--$N$ regime.  In
particular, infrared difficulties of perturbation theory around commuting
matrix configurations should not automatically be identified with an
instability of the large--$N$ limit itself.  Establishing a uniform
extrapolation requires checking finite--$N$ corrections, regulator dependence,
and the order of the infrared, continuum and large--$N$ limits.  The singlet
counting obtained here supplies an analytic benchmark for this program,
but does not settle those dynamical questions.

\paragraph{Emergent geometry and background independence.}

\medskip
\noindent
Finally, the geometric observables used in this work---the extent of space
and the gauge--invariant Gram operators $\Tr(X_aX_b)$---provide a possible
language for studying geometry encoded in matrix degrees of freedom.  Their
singlet counting is not, however, a derivation of a background-independent
quantum-gravitational dynamics.  The present construction specifies a time
direction, a number of matrices and, in the massive theory, deformation
parameters.  Demonstrating how a spacetime geometry could emerge dynamically,
and to what extent the construction can become independent of prescribed
background data, remains a separate question.  Our more limited claim is that
the Gaussian reduction together with the exact gauge-singlet projection offers
a useful, explicitly calculable sector in which to investigate it.

\section{Conclusion}
\subsection{Main results}

\medskip
\noindent
In this paper we studied the singlet-sector structure of mass--deformed
BFSS$_{d+1}$ matrix quantum mechanics from two complementary viewpoints.  The
first is the large--\(d\) Gaussian reduction, in which the interacting matrix model
is replaced by a gauged matrix harmonic oscillator with a self--consistent mass
\(s\).  This captures the bulk Gaussian dynamics and leads to the expected
linear \(d\)-scaling of the leading extent of space.  The second is the
Molien--Weyl projection, which imposes the Gauss law exactly and reorganizes the
physical Hilbert space in terms of gauge--invariant singlet excitations above the
Gaussian vacuum.  The main lesson is that these two descriptions probe different
layers of the same theory: the \(X_a\)-space Gaussian saddle describes the bulk
matrix geometry, while the Molien--Weyl integral describes the holonomy--projected
endpoint or singlet spectrum.

\medskip
\noindent
The central result of Section~\ref{section3} is the emergence of a universal Gram--matrix
counting law in the very--low--temperature Molien--Weyl expansion.  For the
bosonic theory we found
\begin{eqnarray}
Z^{\rm bos}_{N,d}(x)
&=&
1+\frac{d(d+1)}{2}\,x^2+O(x^3),
\qquad
x=e^{-\beta s}.
\end{eqnarray}
For \(N=2\), this coefficient was established both by explicit residue evaluation
of the Molien--Weyl integral and by a direct character calculation.  For \(N>2\),
the same coefficient follows from the character analysis and is therefore expected
to be independent of \(N\).  Its meaning is transparent: it counts the independent
quadratic Gram operators
\(\Tr(X_aX_b)\), with \(a\le b\).  Thus the first nontrivial singlet level is
universal and is controlled by the endpoint Gram structure, even though the full
partition function is generally much richer.

\medskip
\noindent
This also clarifies the precise sense in which BFSS$_2$--like factorization holds.
At very low temperature, the singlet spectrum begins as if it were built out of
\[
k=\frac{d(d+1)}{2}
\]
BFSS$_2$--like quadratic towers.  However, this factorization is not generally an
exact statement about the full Hilbert space.  Starting already at higher \(d\),
new invariant channels appear, such as cubic \(\epsilon\)-type operators, and these
generate additional singlet states not captured by a naive product of BFSS$_2$
partition functions.  Hence the Molien--Weyl projection produces a universal
Gram sector at leading order, but the full singlet Hilbert space contains further
nontrivial invariant structures.

\medskip
\noindent
Section~\ref{section5} gives the complementary Hamiltonian explanation of the exceptional case
where the BFSS$_2$--like factorization becomes exact.  For \(SU(2)\) with two
bosonic matrices, the only independent bosonic singlet creation operators are the
three Gram generators
\begin{eqnarray}
G_{11}=a_{1A}^{\dagger}a_{1A}^{\dagger},
\qquad
G_{22}=a_{2A}^{\dagger}a_{2A}^{\dagger},
\qquad
G_{12}=a_{1A}^{\dagger}a_{2A}^{\dagger}.
\end{eqnarray}
They generate a free polynomial algebra, and all higher bosonic singlets reduce
to products of these Gram operators.  This gives the exact identity
\begin{eqnarray}
Z_{2,3}^{\rm bos}
=
\big(Z_{2,2}^{\rm bos}\big)^3,
\end{eqnarray}
which holds for all temperatures at the special point \((d,N)=(2,2)\).  The
Hamiltonian derivation shows directly why this identity is not a mere infrared
accident: it follows from the minimal invariant tensor structure of the \(SU(2)\)
adjoint representation.

\medskip
\noindent
The same Hamiltonian analysis also explains why the exact factorization does not
extend generically.  For \(d>2\), additional rotational invariant structures become
available.  For \(N>2\), the adjoint invariant tensor algebra becomes much richer,
with higher symmetric traces, higher Casimir tensors, and nontrivial trace
relations.  Thus the equality between simple Clifford or Gram counting and the
full singlet counting is a special \(SU(2)\) phenomenon.  The case \((d,N)=(2,2)\)
therefore occupies a distinguished position: it is the finite--\(N\), two--matrix
realization of an exact BFSS$_2$--factorized singlet Hilbert space.

\medskip
\noindent
A further consequence of this rigidity concerns deconfinement. The exact
factorization at \((d,N)=(2,2)\) implies the same singlet thermodynamics as
three BFSS$_2$ copies, but does not exclude a smooth deconfinement--like
rearrangement of the holonomy distribution. There is no genuine Hagedorn
or deconfinement phase transition at fixed finite \(N\). More generally,
\(d=1\) is exceptional because the large--\(N\) holonomy instability is
absent, while \(N=2\) is exceptional for all \(d\). At \(N=2\), all Polyakov
moments are constrained functions of a single holonomy angle and cannot
act as independent instability modes. Nevertheless, a smooth
deconfinement--like holonomy crossover remains possible at finite \(N\),
including at \((d,N)=(2,2)\), as demonstrated above.
In the strict \(N\to\infty\) limit, a genuine nonanalytic
deconfinement/Hagedorn transition can emerge for \(d>1\).

\medskip
\noindent
The supersymmetric analysis of Section~\ref{section4} shows that the same Gram logic survives
after fermions are included.  The leading low--temperature singlet spectrum is then
organized by the quadratic operators
\[
\Tr(X_aX_b),\qquad
\Tr(X_a\psi_r),\qquad
\Tr(\psi_r\psi_s),
\]
leading to the universal expansion
\begin{eqnarray}
Z^{(d;n_f)}_{SU(N)}(x_b,x_f)
&=&
1
+\frac{d(d+1)}{2}\,x_b^2
+d\,n_f\,x_bx_f
+\frac{n_f(n_f-1)}{2}\,x_f^2
+O\!\Big((x_b,x_f)^3\Big).
\end{eqnarray}
Thus the infrared supersymmetric singlet spectrum is governed by the Gram sector
and its fermionic extensions, while higher singlet structures enter only at higher
orders.

\medskip
\noindent

Finally, Section~\ref{section6} outlines how these analytic results
can be used in Monte Carlo studies, with further
numerical details provided in the original arXiv
version of this work~\cite{Ydri:2026MolienV1}.  The Gaussian BFSS/BMN system admits both an
\(X_a\)-space matrix harmonic oscillator representation and a \(\theta_i\)-space
Molien--Weyl representation.  Their equivalence, up to vacuum--energy
normalizations and possible fermionic condensates, provides a useful benchmark for
HMC/RHMC simulations of the full theory.  The comparison between full,
bosonic, Gaussian Molien--Weyl, and bosonic Molien--Weyl descriptions gives a way
to separate from each other bulk commutator dynamics, holonomy projection effects, singlet
counting, and fermionic holonomy contributions.

\subsection{Summary}

\noindent
The main conclusions are the universal quadratic Gram-singlet counting law,
the exceptional exact BFSS$_2$ factorization at $(d,N)=(2,2)$, and the
separation between Gaussian bulk observables and holonomy-projected singlet
excitations.  The $N=2$ residue calculations provide precise benchmarks for
this separation, while the supersymmetric extension identifies the analogous
boson--fermion quadratic singlets.  As emphasized in Section~\ref{section6},
controlling infrared flat directions in the undeformed interacting theory,
understanding the genuinely nonperturbative large--$N$ limit, and relating
matrix invariants to background-independent geometry remain important future
steps rather than results claimed here.

\section{Acknowledgments}

The author acknowledges the role of ChatGPT-5.5 and earlier versions as an AI collaborator in discussing, testing, and refining scientific ideas, as well as in language editing, LaTeX assistance, symbolic calculations, and literature searches.

\end{document}